\documentclass[twocolumn,times]{aastex63}

\received{2020~$November$~19}
\revised{2021~$June$~11}
\accepted{2021~$June$~14}
\shorttitle{Multiple Mg\,{\sc ii} Absorption Systems to Quadruply Lensed Quasar H1413+1143}
\shortauthors{Okoshi et al.}
\begin{document}

\title{Multiple Mg\,{\sc ii} Absorption Systems in the Lines of Sight to Quadruply Lensed Quasar H1413+1143}

\correspondingauthor{Katsuya Okoshi}
\email{okoshi@rs.tus.ac.jp}

\author[0000-0003-3466-3876]{Katsuya Okoshi}
\affiliation{Institute of Arts and Sciences, Tokyo University of Science, 6-3-1, Niijyuku, Katsushika, Tokyo 125-8585, Japan:okoshi@rs.tus.ac.jp}
\author[0000-0002-3245-4758]{Yosuke Minowa}
\affiliation{Subaru Telescope, National Astronomical Observatory of Japan, 650 North A'ohoku Place, Hilo, Hawaii 96720, U.S.A}
\author[0000-0003-3954-4219]{Nobunari Kashikawa}
\affiliation{Department of Astronomy, The University of Tokyo, 7-3-1 Hongo, Bunkyo-ku, Tokyo 113-0033, Japan}
\author[0000-0002-5464-9943]{Toru Misawa}
\affiliation{School of General Education, Shinshu University, 3-1-1 Asahi, Matsumoto, Nagano 390-8621, Japan}
\author[0000-0001-9044-1747]{Daichi Kashino}
\affiliation{Institute for Advanced Research, Nagoya University, Furocho, Chikusa-ku, Nagoya, Aichi 464-8602, Japan}
\author[0000-0001-6501-3871]{Hajime Sugai}
\affiliation{Environment and Energy Department, Japan Weather Association, Sunshine 60 Bldg. 55F, 3-1-1 Higashi-Ikebukuro, Toshima-ku, Tokyo 170-6055, Japan}
\author[0000-0001-6473-5100]{Kazuya Matsubayashi}
\affiliation{Okayama Observatory, Kyoto University, 3037-5 Honjo, Kamogata-chi, Asakuchi, Okayama 719-0232, Japan}
\author{Atsushi Shimono}
\affiliation{Graduate School of Media and Governance, Keio University, 5322 Endo, Fujisawa, Kanagawa, 252-0882, Japan}
\author[0000-0002-5443-0300]{Shinobu Ozaki}
\affiliation{National Astronomical Observatory of Japan, Mitaka, Tokyo 181-8588, Japan}

\begin{abstract}

We find multiple Mg\,{\sc ii} absorption systems at redshift $z=1.66$, $2.069$, and $2.097$ in the spatially resolved spectra of the quadruply gravitationally lensed quasar H1413+1143 utilizing the Kyoto tridimensional spectrograph \,{\sc ii} (Kyoto 3D\,{\sc ii}) spectrograph on board the Subaru telescope. 
Here we present the first measurement of differences in Mg\,{\sc ii} absorption strength of the multiple intervening absorbers, which include  ones identified as damped Lyman $\alpha$ (DLA) absorption systems.  
Our detection of the significant Mg\,{\sc ii} absorptions in the spatially resolved spectra reveals the inhomogeneous chemical enrichment on scales of about $12$ kpc within the separation of the four sightlines. 
For the DLA system at $z =1.66$, the rest equivalent widths of the Mg\,{\sc ii} absorption lines between the four spatially resolved lines of sight change by factors of up to $\sim 6$, which trace the variations in the H\,{\sc i} absorption strength. 
This suggests that inhomogeneous cold absorbers that give rise to the strong H\,{\sc i}/Mg\,{\sc ii} absorptions dwell on a scale of about $6-12$ kpc between the four lines of sight. 
We also investigate the degree of variation in the equivalent width of the absorption lines between the lines of sight. 
We find that the systems giving rise to strong absorptions in the spectra of the quadruply lensed quasars tend to have 
a high degree of variation in absorption strength between the lines of sight toward the lensed quasars. 

\end{abstract}



\section{Introduction} \label{sec:intro}

The absorption line systems found in quasar spectra, the so-called `quasar absorption systems', provide us with a unique probe of galaxy formation and evolution processes. 
The systems offer valuable opportunities to explore the physical and chemical conditions of intergalactic media and/or intervening galaxies (e.g., the amounts of neutral gas and metals, the
gas dynamics). 
In particular, metal absorption systems have been studied extensively to place stringent constraints on the
galaxy formation and evolution processes 
because the absorption feature provides the basic information about the abundance of absorbing gas and its metal content. 

The gravitationally lensed quasars have been given much attention, since the quasars show the multiple spectra toward the background images, which provides us with valuable information on the intervening absorption systems.  
The variations of dynamical and chemical structures in the intervening absorbers produce a variety of absorption features in the spectra, such as the absorption strength and the velocity width of the absorption lines. 
Focusing on triply or quadruply lensed quasars, several quasars have multiple spectra of the separate images where hydrogen and/or  metal absorptions at the same redshift are identified at the same redshift (e.g., APM08279+5255, J1004+4112). 
The multiple absorption line features is a unique probe of the detailed spatial structures of the absorbing gas and metals on scales of the separation of the sightlines (e.g., tens of parsecs to a few tens of kiloparsecs).

The quasar H1413+1143 at emission redshift $z_{\rm em} = 2.54$ is one of the quasars given much attention due to a quadruple gravitational lens, which is often referred to as the Cloverleaf \citep[e.g.][]{Hazard84, Drew84, Magain88}. 
The spectrum of quasar H1413+1143 shows a variety of absorption lines together with a presence of the broad absorption lines (BALs). 
In the previous studies, however, the images separated by the gravitationally lensing were not available due to low spatial resolution. 
The spectroscopy contained all the four absorbing components in a {\it composite spectrum} that is {\it not} spatially-resolved . 
In the composite spectrum, \citet{Hazard84} and \citet{Drew84} identified the BALs and metal absorption lines (e.g., Al\,{\sc iii}, C\,{\sc iv}) at absorption redshifts $z_{\rm abs} = 1.660$ and $2.068$. 
\citet{Turnshek88} subsequently presented a detailed study of the
spectrum including the BALs and metal absorption lines at $z_{\rm abs} \sim 1.66, 1.87$, and $2.07$.
Utilizing the New Technology Telescope, \citet{Magain88} first identified quasar H1413+1143 as a gravitationally lensed system where four image-components all lie within about $0.7$ arcsec of the image center. 
Furthermore, the medium-resolution Faint Object Spectrograph (FOS) spectra on board the Hubble space telescope (HST) successfully separated the four individual images of quasar H1413+1143. 
In the multiple spectra of the four individual images, the intervening absorbers give rise to {\it neutral hydrogen absorption lines} at $1.7 < z_{\rm abs} < 2.5$ including a damped Lyman $\alpha$ (DLA) system (the H\,{\sc i} column densities $N_{\rm HI}$ $>$ $10^{20.3}$ cm$^{-2}$ ) at $z_{\rm abs} = 1.6595$ and a sub-DLA ($10^{19}$ cm$^{-2}$ $< N_{\rm HI} < 10^{20.3}$ cm$^{-2}$) at $z_{\rm abs} =2.0969$ \citep{Monier98}. 
The subsequent study of absorbers in the near-UV spectra, utilizing the Space Telescope Imaging Spectrograph (STIS) on board the HST, 
revealed (sub-)DLA systems at $z_{\rm abs} = 1.44$ and
$1.49$ in addition to the DLA at $z_{\rm abs} \sim 1.66$ \citep{Monier09}. 
The multiple lines of sight toward the quadruply lensed quasar can probe the properties of the intervening absorption systems not only along the sightlines but along the transverse direction toward the sightlines. 
Over the sightline separations about within $1.4$ arcsec (corresponding to the transverse size of about $12$ kpc at $z \sim 1.66$), the H\,{\sc i} column densities in the three (sub-)DLA systems typically change by factors of $2-20$ \citep{Monier09}, indicating 
a presence of the variations in H\,{\sc i} gas density on the scales within ten kilo-parsecs of the sightline separations.

In the spatially resolved  spectra toward four image-components (A, B, C, and D), 
the variation of {\it metal} absorption strengths in the four spectra has also been given much attention. 
\citet{Monier98} found a strong 
Fe\,{\sc ii} absorption system at $z_{\rm abs} \sim 1.44$ in the HST-FOS spectrum toward component B,  
while the Fe\,{\sc ii} absorption line is absent toward the other C component. 
Furthermore, the authors also presented a detailed study of metal absorption
systems at $z_{\rm abs} = 0.6089, 1.3547, 1.4377, 1.6595, 2.0680$, and $2.0969$, which show the presence of various metal
absorptions, e.g., Si\,{\sc ii} $\lambda1526$, Si\,{\sc iv} $\lambda \lambda 1393, 1402$, C\,{\sc ii} $\lambda 1334$, 
C\,{\sc iv}  $\lambda \lambda 1548, 1550$, Al\,{\sc ii}  $\lambda 1670$, Al\,{\sc iii}  $\lambda\lambda 1854, 1862$, and 
several Fe\,{\sc ii} transitions {\it except for Mg\,{\sc ii} $\lambda \lambda 2796, 2803$}. 
Similar to the H\,{\sc i} absorption systems, the metal absorption systems provide clear evidence
for differences in absorption strength in the four individual spectra. 
Specifically, a significantly strong absorption is detected in a spectrum toward one of the lensed image-components but the others. 
This suggests that chemical enrichment differs at least on scales within the separations 
of the sightlines similarly to the variation in H\,{\sc i} absorption strength.
The results are based mainly on the metal absorption lines due to the ion-transitions within  
the wavelength coverage of the HST-FOS spectra. 
However, the studies have not been able to measure metal transition lines like the Mg\,{\sc ii} $\lambda \lambda 2796, 2803$ doublet lines.

The Mg\,{\sc ii} absorption system is one of the most studied tracers of chemical enrichment 
in cold gas since the ionization potential ($\sim 1.1$ Ryd)  is close to that of neutral hydrogen \citep[e.g.][]{Bergeron86}.
Furthermore, the Mg\,{\sc ii} absorption is ideal for investigating metal-enriched 
cold gas, which is optically thick against the photo ionizing radiation, where it 
arises in H\,{\sc i} absorption systems with a wide range of H\,{\sc i} column densities, $N_{\rm HI}$ $\sim$ $10^{16} - 10^{22}$ cm$^{-2}$ 
\citep[e.g.][]{Steidel92, Churchill00, Rigby02, Rao06, Rao17, Menard09}. 
For example, DLA systems often give rise to strong Mg\,{\sc ii} absorption lines.  The number fraction of Mg\,{\sc ii} systems that are also identified as (sub-)DLA systems increases toward large equivalent widths of the Mg\,{\sc ii} absorption lines. 
For 70 DLA systems at $z_{\rm abs} < 1.7$, $40 \%$ or more Mg\,{\sc ii} systems with 
the rest equivalent width $W_{\rm \lambda} \ga 2.0$ ${\rm \AA}$ are also identified as DLA systems \citep{Rao17}.  
In addition, at high redshift $z>2$, there is also a trend that Mg\,{\sc ii} systems with 
$W_{\rm \lambda} \sim 0.3$ ${\rm \AA}$  ($\sim 40 \%$) are likely associated with DLA systems \citep[e.g.][]{Matejek13}. 
The results suggest that strong Mg\,{\sc ii} systems originate primarily from gas systems with high neutral hydrogen column densities in the circumgalactic medium (CGM) around galaxies, particularly at high redshift. 
Previous studies revealed that Mg\,{\sc ii} systems trace the dynamical and chemical structure 
of the CGM such as gas-flows \citep[e.g.][]{Tremonti07, Rubin10, Kacprzak11, Nestor11, Bouche12, Martin12, Martin19, Bordoloi14, Rubin14, Rubin18a, Schroetter16, Schroetter19}.
Constraining the properties of CGM through the Mg\,{\sc ii} absorption has been based {\it statistically} on the studies using a Mg\,{\sc ii} absorption line in a spectrum along a line of {\it single} sight toward a background quasar or galaxy.  
Utilizing the single Mg\,{\sc ii} absorptions, there has been growing {\it statistical} evidence for metal enrichment in the CGM.

Here we focus on metal distribution based on {\it the spatial extent of individual Mg\,{\sc ii} absorption lines} in the spectra toward multiple background {\it quasars} (e.g., quadruply or triply lensed quasars).      
The current sample of multiple Mg\,{\sc ii} absorption systems is quite limited, since there are only  a handful 
of  quadruply or triply lensed quasars giving rise to Mg\,{\sc ii} absorption lines in the spectra with high resolution \citep[e.g.][]{Rauch02, Ellison04, Chen14, Rubin18b}. 
For example, \citet{Rubin18b} focused on a quadruply lensed quasar J014710+463040 at $z_{\rm em}=2.377$. 
The authors measured Mg\,{\sc ii} absorption strengths in the four separate spectra toward the quadruply lensed images 
and investigated a degree of spatial coherence for the multiple Mg\,{\sc ii} absorbers over scales of up to $\sim 20$ kpc at $z_{\rm abs}$ $<1$.
At $z_{\rm abs}$ $>1$, quadruply or triply lensed quasars (e.g., J1004+4112, APM08279+5255) provide the multiple Mg\,{\sc ii} absorption lines \citep[e.g.][]{Oguri04, Ellison04}. 
However, there is still a quite limited sample to evaluate the variation in Mg\,{\sc ii} absorption strength due to small incidence rates of 
the absorption lines and/or small separations of the sightlines. 
For the quadruply lensed quasar H1413+1143, there have not been identified Mg\,{\sc ii} absorption lines individually in the four spatially resolved spectra.

Here, utilizing the Kyoto tridimensional spectrograph \,{\sc ii} (Kyoto 3D\,{\sc ii}) spectrograph on board the Subaru telescope, 
we find multiple Mg\,{\sc ii} absorption systems at $z_{\rm abs} = 1.66$, $2.069$, and $2.097$ with high incidence rates on scales of up to $\sim12$ kpc in the spatially resolved spectra of quasar H1413+1143. 
The high signal-to-noise (S/N) spectroscopy covers the wavelengths of $7300$ ${\rm \AA}$ $< \lambda < 9200$ ${\rm \AA}$ 
and also achieved the S/N necessary to assess the Mg\,{\sc ii} absorption equivalent widths 
associated with the systems previously identified in hydrogen and the other metal absorptions. 
The measurements of equivalent widths in the separate components for the Mg\,{\sc ii} $\lambda \lambda 2796, 2803$ doublet absorption lines at $z>1$ offer a probe of a presence of variations in Mg\,{\sc ii} absorption strength like that of H\,{\sc i} absorption lines  
due to the transition with the ionization energy similar to that of Mg\,{\sc ii}. 
For the Mg\,{\sc ii} absorption systems in the spectra of the quadruply lensed quasar H1413+1143, we here focus on the followings. 
Does the variation in Mg\,{\sc ii} absorption strength between the multiple lines of sight  trace that of neutral hydrogen with the H\,{\sc i} column densities changing by factors of up to $\sim 20$ at $z_{\rm abs} \sim 1.66$ \citep[e.g.][]{Monier09}? 
Is the variation in Mg\,{\sc ii} absorption strength consistent with those in high-/low-ion absorption strength? 
Focusing on the Mg\,{\sc ii} absorption lines in the gravitationally lensed quasar, does the variation in absorption 
strength depend on separations between the absorbers or equivalent widths of the absorbers?

In Section 2, we describe our observations in detail. The results are presented in Section 3. Several
implications from our results are derived in Section 4. Finally, we draw our conclusions in Section 5. 
In this paper, we adopt a standard $\Lambda$CDM cosmological model with parameters, 
$\Omega_{\rm 0} = 0.3$, $\Omega_{\rm \Lambda} = 0.7$, and the Hubble constant $H_{\rm 0} = 70$ km s$^{-1}$ Mpc$^{-1}$.

%%%%%%%%%%%%%%%%%%%%%%%%%%%%%%%%%%%%%%%

\section{Observations}

\setcounter{figure}{0}

\begin{figure}[h]
\begin{center}
%\includegraphics[width=10cm]{/Users/okosh/Documents/Latex/paper8/H1413_image.pdf}
%[two column]\includegraphics[width=10cm]{fig1.pdf}
\includegraphics[width=8cm]{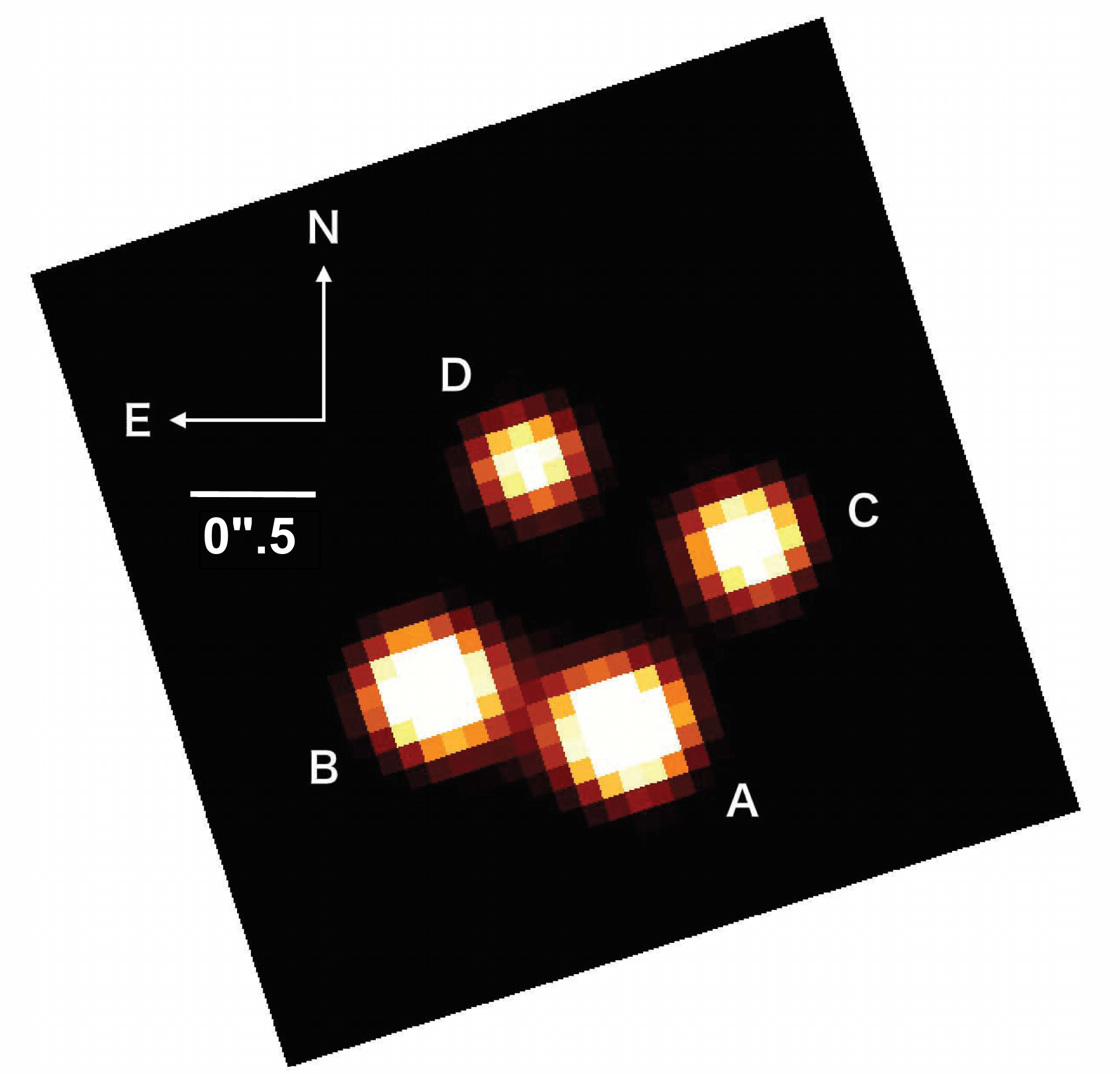}
\end{center}
\caption{Spatially resolved image of four components A/B/C/D in the quadruply gravitationally lensed quasar H1413+1143 on the basis of the high spatial resolution and high-S/N spectroscopy with an optical multimode spectrograph, the Kyoto tridimensional spectrograph \,{\sc ii} (Kyoto 3D\,{\sc ii}). 
The image was generated by averaging the data cube in a wavelength range between $7300$ and $9200$ ${\rm \AA}$. We adjusted the position of the objects in each cube so as to compensate for the position shift due to the atmospheric dispersion.
\label{H1413_image}}
\end{figure}

The quasar H1413+1143 at $z_{\rm em}=2.54$ has four image components A/B/C/D by gravitational lensing which all lie within $0.7$ arcsec of the center (Figure $\ref{H1413_image}$). 
We observed the H1413+1143 on 2017 February 8 with the Kyoto 3D\,{\sc ii} optical integral field units (IFU) at a Nasmyth focus of the Subaru telescope. 
In combination with a 188-element adaptive optics (AO) system, Subaru AO188 \citep{Hayano08, Minowa10}, 
the Kyoto 3D\,{\sc ii} can provide the AO-assisted optical integral-field spectroscopy with $\lambda > 6400$ ${\rm \AA}$ \citep{Matsubayashi16}.
The Kyoto 3D\,{\sc ii} is installed with a Hamamatsu fully depleted charge-coupled device (CCD; \citet{Mitsuda16}). 
The observed wavelength band was No.5  \citep{Matsubayashi16} which can cover the range of the wavelength $7300$ ${\rm \AA} < \lambda < 9200$ ${\rm \AA}$ (a wavelength sampling rate of $3.8$ ${\rm \AA}$ pixel$^{-1}$) with high sensitivity at $\sim 9000 {\rm \AA}$. 
The spectral resolution is $R \sim 1200$. 
The field of view (FoV) is about $3^{\prime \prime}.21 \times 2^{\prime \prime}.52$. 

%fig.1

We used a laser guide star AO (LGS-AO) mode of AO188 \citep{Hayano08, Hayano10}.
Utilizing component A ($V = 17.9$ mag) of the H1413+1143 image as a tip-tilt guide star (TTGS), 
we obtained the AO-corrected FWHM of $0.42$ arcsec. 
The total on-source exposure time is  $12,000$ sec ($1200$ sec $\times 10$ frames). 

We used the custom-made IRAF scripts \citep{Sugai10} which are adapted for the 
installed deep depletion CCD.  
The bias subtraction, the spectrum extraction, and the flat-fielding are included in the reduction process.  
We used the L.A.Cosmic for removing cosmic rays \citep{vanDokkum01}. 
After the wavelength calibration, sky subtraction was performed on the sky aperture spectra. 
Based on the Kyoto 3D\,{\sc ii} spectroscopy,  
we simultaneously obtained the spectra of the object and sky, which is about $29^{\prime \prime}$ away from the object field. 
The flux was calibrated using a standard star. 
Here, using the normalized spectrum of the standard star Feige 67, we corrected the atmospheric absorption features in the spectra. 
After the flux calibration, we combined the $10$ frames and carried out the spectral fitting in the obtained spectra. 
The individual spectrum was extracted from the data cube for each lensed object. We used an aperture size of $0.84$ arcsec around the center of each object. 
The continuum fitting was performed by fitting a third-order polynomial function to the spectrum for each object in a wavelength range between $7350$ and $7700$ ${\rm \AA}$, except the range where the absorption lines are located.
%%%%%%%%%%%%%%%%%%%%%%%%%%%%%%%%%%%%%%%%%%%%%%%%%%%

\section{Results}

\setcounter{table}{0}

\begin{figure*}[t]
\begin{center}
\includegraphics[width=15cm]{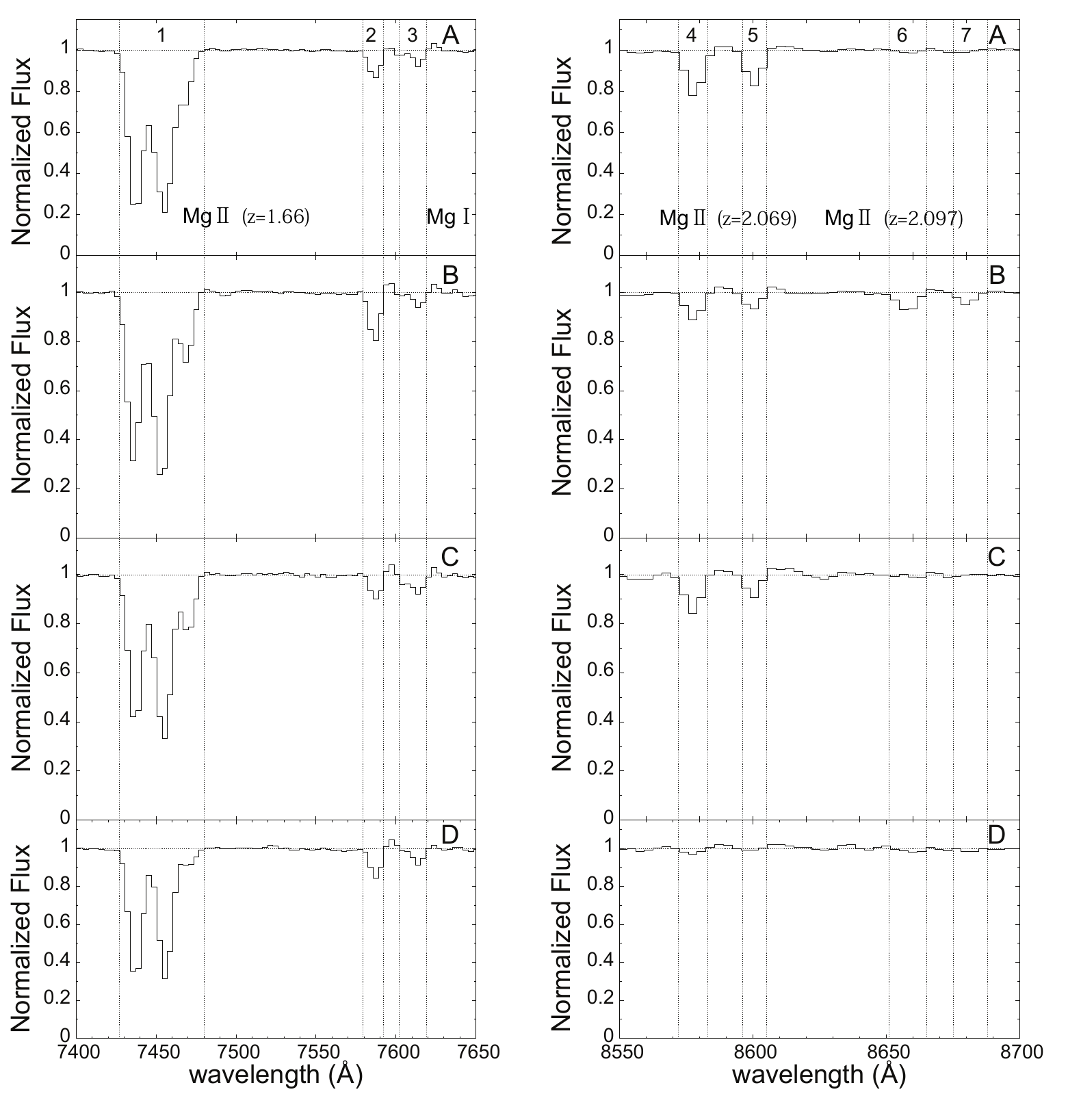}
\end{center}
\caption{Individual spectra (Nos.1-7) in the four sightlines toward the A/B/C/D components of the H1413+1143 images (top to bottom).  
Left: Mg\,{\sc ii} and Mg\,{\sc i} absorption lines at $z_{\rm abs}=1.66$. Right: Mg\,{\sc ii} absorption lines at $z_{\rm abs}=2.069$ and $2.097$. 
\label{spec_figure_new}}
\end{figure*}

In Figure $\ref{spec_figure_new}$, we show the extracted spectra within the limited wavelength ranges around detected absorption lines:  
$7400$ ${\rm \AA} < \lambda <7650$ ${\rm \AA}$ (left panel) and $8550$ ${\rm \AA} < \lambda < 8700$ ${\rm \AA}$ (right panel). 
In the spectra, we find seven absorption features (Nos.1-7) in the lines of sight toward components A/B/C/D of the gravitationally lensed image (hereafter, the A/B/C/D spectra, respectively). 
The spectra shown in the left panel of Figure $\ref{spec_figure_new}$ clearly present a strong Mg\,{\sc ii} absorption line at  $z_{\rm abs} \sim 1.66$. 
The absorption line No.1 appears to suffer from a self-blending effect; blue and red lines of the doublet caused by two transitions, $\lambda \lambda 2796, 2803$, are blended due to the spectral resolution ($R \sim 1200$). 
We find that the line is well fitted by a two-component Gaussian profile with different central wavelengths and widths; the bluer and stronger component No.1a and the other No.1b. 
To fit the two components simultaneously, we adopt two double-Gaussian functions (for the rest wavelengths of $2796$ and $2803$ ${\rm \AA}$), for which two redshifts are treated as free parameters.  The relative central wavelengths of the Mg\,{\sc ii} double lines are fixed.  Two independent velocity widths are assumed for the two components at different redshifts.
The amplitudes of all the four lines are treated as free parameters. 
The best-fit parameters of the profiles are determined using a $\chi^{2}$ minimization method.
The resultant errors account for the effect of degeneracy between the parameters for fitting the line by the two components.

\begin{figure}[h]
\begin{center}
%\includegraphics[width=10cm]{/Users/okosh/Documents/Latex/paper8/spec_fitting.pdf}
%[two column]\includegraphics[width=10cm]{fig3.pdf}
\includegraphics[width=9.5cm]{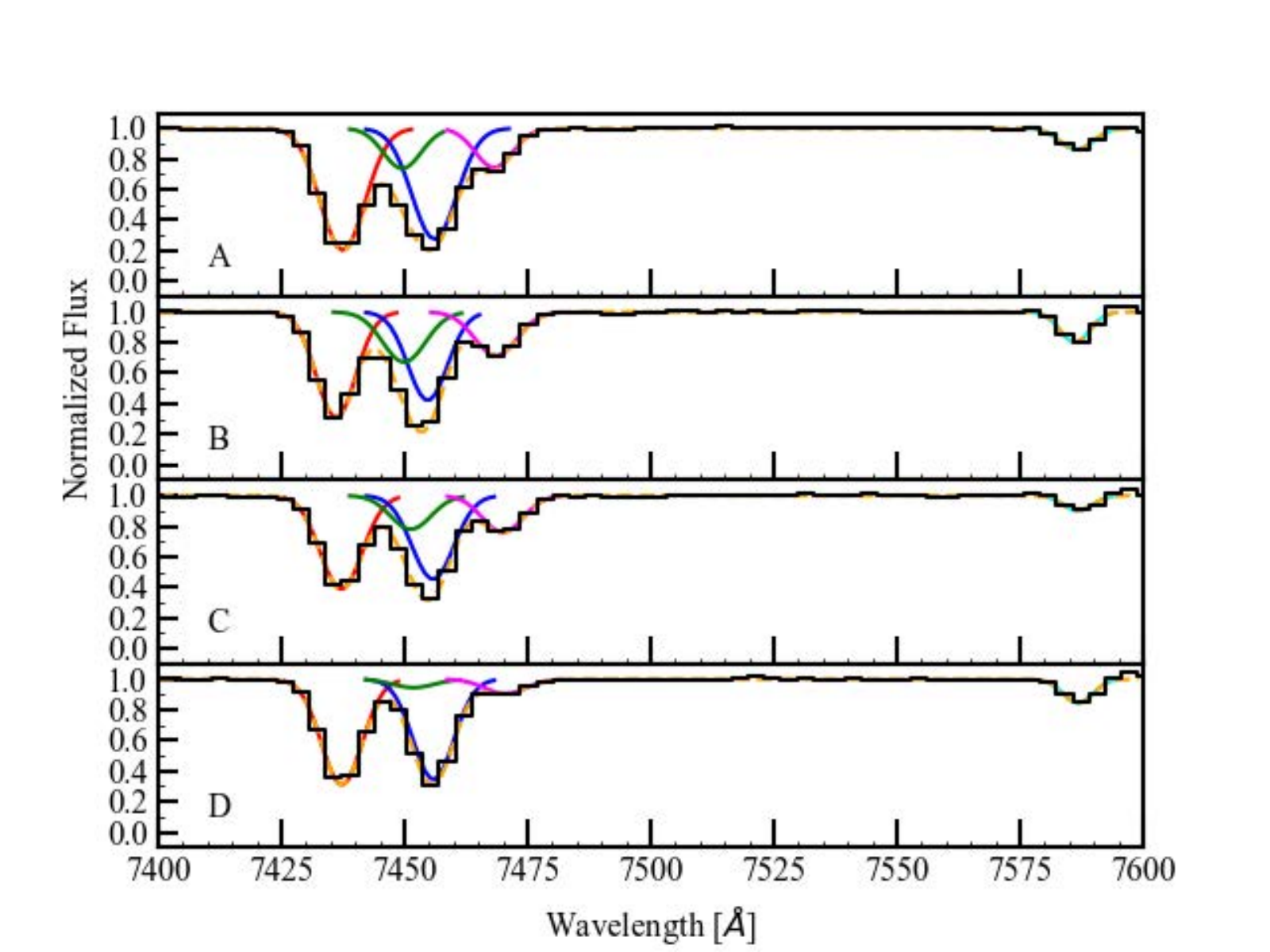}
\end{center}
\caption{Individual spectra in the four sightlines toward the A/B/C/D images (top to bottom), where the absorption line (No.1) at $z_{\rm abs} = 1.66$  (orange) are fitted by Gaussian profiles of two components; one (No.1a) at $z_{\rm abs}=1.660$ with a doublet at $\lambda=2796$ (red) and  $2803$ (blue) ${\rm \AA}$ and another (No.1b) at $z_{\rm abs}=1.664$ with a doublet at $\lambda=2796$ (green) and  $2803$ (magenta) ${\rm \AA}$ .  
\label{spec_fitting}}
\end{figure}

In Figure $\ref{spec_fitting}$, we present the A/B/C/D spectra in which the absorption line (No.1) is fitted by the doublet lines of the two components: one (No.1a) at $z_{\rm abs}=1.660$ giving rise to a doublet at $\lambda=2796$ (red) and  $2803$ (blue) ${\rm \AA}$ in the rest frame and another (No.1b) at $z_{\rm abs}=1.664$ with a doublet at $\lambda=2796$ (green) and  $2803$ (magenta) ${\rm \AA}$.  
In Table $\ref{table_EW}$, we present both the {\it total} rest equivalent widths of the blended absorption line (No.1), $W_{\rm MgII}^{0} (\lambda 2796)+W_{\rm MgII}^{0} (\lambda 2803)$, and 
the rest equivalent widths of the doublets for the two components (No.1a and 1b) in the four A/B/C/D spectra. 
In addition to the Mg\,{\sc ii} absorption lines, two Mg\,{\sc i} $\lambda 2853$ absorption lines are identified 
at $z_{\rm abs}=1.659$ and $1.667$ (Nos.2-3). 
\setcounter{table}{0}
The rest equivalent widths are also presented in Table $\ref{table_EW}$.  
Furthermore, we identify a Mg\,{\sc ii} absorption doublet (Nos.4-5) at $z_{\rm abs}=2.069$ in the A/B/C spectra and one (Nos.6-7) at $z_{\rm abs}=2.097$ in the B spectrum.  

% Fig 2
%%[two column]\begin{figure}[h] -> figure* & [t]

% Fig 3

In summary, the four spectra toward the spatially resolved four images present the Mg\,{\sc ii} doublet line ($\lambda \lambda 2796, 2803$), which is composed of two components are identified at absorption redshift $z_{\rm abs} \sim 1.66$ (No.1), the Mg\,{\sc ii} doublet lines at $z_{\rm abs}$ $=$ $2.069$ (Nos.4-5) and $2.097$ (Nos.6-7), and the two Mg\,{\sc i} absorption lines at $z_{\rm abs} = 1.659$ (No.2) and $1.667$ (No.3). 
In Table $\ref{table_EW}$, the rest equivalent widths of the absorption lines are summarized.  
For undetected absorption lines in the spectra, we include upper limits on the rest equivalent widths based on our detection limit.

The presence of the multiple Mg\,{\sc ii}/Mg\,{\sc i} absorption lines in the spectra toward the {\it separate} four-images are reported for the 
first time while a composite Mg\,{\sc ii} absorption line at $z_{\rm abs} \sim 1.66$ is previously identified in a spectrum in the {\it unseparated} line of sight toward the {\it composite} image using  the Keck Low Resolution Imaging Spectrometer (LRIS) \citep[e.g.][]{Drew84}.  
The HST-FOS provided the separate spectra in the lines of sight toward the resolved  images.  
However, the FOS does not extend the wavelengths of Mg\,{\sc ii} absorption lines at $z_{\rm abs} \sim 1.66$ in the A/B/C/D spectra (e.g., Monier et al. 1998; Table 3 in Monier et al. 2009).
The wavelength coverage of the Kyoto 3D\,{\sc ii} spectra allows the first measurement of equivalent widths in the separate spectra 
for the Mg\,{\sc ii} absorption line at $z_{\rm abs} \sim 1.66$ together with the Mg\,{\sc i} absorption lines and other Mg\,{\sc ii} absorption lines at $z_{\rm abs}=2.069$ and $2.097$.

The gravitationally lensed images provide the multiple lines of sight that offer a unique and valuable probe of the spatial structure of the intervening absorbers by mapping the transverse dimension. 
Here we compute the physical distance in the transverse direction to use the following equations \citep[e.g.][]{Smette92,Cooke10}. 
The physical distance, $d$, in the transverse direction between different lines of sight depends on the redshifts of the absorber $z_{\rm abs}$ and the lens $z_{\rm lens}$. 
In the case that  $z_{\rm abs} < z_{\rm lens}$, $d = \theta_{\rm obs} D_{\rm OA}$ where $\theta_{\rm obs}$ is the angular separation and $D_{\rm OA}$ is the angular diameter distance from the observer to the absorber. 
When $z_{\rm abs} > z_{\rm lens}$, $d=\theta_{\rm obs}D_{\rm OL}D_{\rm SA}/D_{\rm SL}$ where $D_{\rm OL}$, $D_{\rm SA}$, and $D_{\rm SL}$ are the angular diameter distances from the observer to the lens, the source to the absorber, and the source to the lens, respectively. 
For the absorbers in lines of sight to the lensed quasar H1413+1143 at source redshift $z_{\rm source}$ $=2.54$,  
the lens object has not been precisely identified.
The measurements of the time delays between the lensed images provide estimations of possible redshifts of the lensing galaxy, $z_{\rm lens}$ $\sim 1.88-1.95$ \citep[e.g.][]{Goicoechea10, Akhunov17}. 
Here we adopt the redshift of lens $z_{\rm lens}=1.90$.  
The observed angular separations between the four spatially resolved lines of sight A/B/C/D are 
($\theta_{\rm obs}$($\overline{\rm AB}$), $\theta_{\rm obs}$($\overline{\rm AC}$), $\theta_{\rm obs}$($\overline{\rm AD}$), $\theta_{\rm obs}$($\overline{\rm BC}$), $\theta_{\rm obs}$($\overline{\rm BD}$), $\theta_{\rm obs}$($\overline{\rm CD}$)) = ($0.753$, $0.872$, $1.118$, $1.359$, $0.967$, $0.893$) arcsec \citep[e.g.][]{Monier09}. 
For the system at $z_{\rm abs}=1.66$, the angular separations give the proper distance in the transverse direction between the four spatially resolved lines of sight A/B/C/D; 
($d$($\overline{\rm AB}$), $d$($\overline{\rm AC}$), $d$($\overline{\rm AD}$), $d$($\overline{\rm BC}$), $d$($\overline{\rm BD}$), $d$($\overline{\rm CD}$)) = ($6.4$, $7.4$, $9.5$, $11.5$, $8.2$, $7.6$) kpc.
For the system at $z_{\rm abs}=2.069$, 
($d$($\overline{\rm AB}$), $d$($\overline{\rm AC}$), $d$($\overline{\rm AD}$), $d$($\overline{\rm BC}$), $d$($\overline{\rm BD}$), $d$($\overline{\rm CD}$)) = ($4.2$, $4.9$, $6.3$, $7.6$, $5.4$, $5.0$) kpc.
It should be noted that the uncertainty in the redshift of the lens may affect the estimation of the physical distance in the transverse direction between the lines of sight at $z_{\rm abs}$ $>$ $z_{\rm lens}$. 
The possible lensing galaxy may belong to a galaxy cluster/group, which implies the lens redshift $z_{\rm lens}$ $\sim$ $1.7$ \citep[e.g.][]{Kneib98a, Kneib98b, MacLeod09}.
If the lens redshift $z_{\rm lens} = 1.7$ is adopted, for the systems with $z_{\rm abs}$ $>$ $z_{\rm lens}$, the distance would decrease by a factor of $0.68$. 
For the system at $z_{\rm abs}=2.069$, the largest distance between the line of sight to image B and that to image C is $\sim 5.2$ kpc, and the smallest one between the lines of sight to images A and B is $\sim 2.9$ kpc.

\subsection{Mg\,{\sc ii}/Mg\,{\sc i} System at $z_{\rm abs} = 1.66$}

The spectra shown in the left panels (from top to bottom) of Figure $\ref{spec_figure_new}$ clearly present a strong Mg\,{\sc ii} absorption line (No.1) at  $z_{\rm abs} \sim 1.66$ in all four A/B/C/D spectra. 
In Table $\ref{table_EW}$, we present the rest equivalent widths of the Mg\,{\sc ii} absorption line.  
The rest equivalent width in the A spectrum is the largest in the four A-D spectra whereas that in the D spectrum is the smallest. 
Comparing the rest equivalent widths in the four spectra to each other, the rest equivalent widths change as A $>$ B $>$ C $>$ D by factors of up to $\sim 1.4$. 
Due to the blending effect on line No.1, two components (No.1a $\&$ 1b) are needed to fit the line profile. 
In Table $\ref{table_EW}$, the rest equivalent widths of the Mg\,{\sc ii} $\lambda \lambda 2796, 2803$ doublet absorption lines in the two components (No.1a $\&$1b) are also presented separately. 
For the No.1a component, the rest equivalent widths are clearly larger than those in the No.1b component. 
For example, line No.1a in the A spectrum has a rest equivalent widths of $\sim 3.0~{\rm \AA}$, which is about three times larger than those of line No.1b. 
The rest equivalent width of line No.1a in the A spectrum is the largest in the A-D spectra whereas that in the C spectrum is smallest. 
The rest equivalent widths changes by factors of up to $\sim 1.5$. 
For example, the rest equivalent width of the Mg\,{\sc ii} $\lambda 2796$ absorption line, $W_{\rm MgII}^{0} (\lambda 2796)$, changes as A $>$ D $>$ B $>$ C. 
The second component of No.1b shows moderate absorption strengths with rest equivalent widths $\sim 1.0$ ${\rm \AA}$ or less. 
By contrast to line No.1a, the rest equivalent widths of line No.1b clearly change by factors of up to about $6$: B $>$ A $>$ C $>$ D. 
The No.1b line at $z_{\rm abs}=1.664$ shows large differences in equivalent width between the lines of sight rather than the No.1a line at $z_{\rm abs}=1.660$. 
The result indicates a presence of the variation in Mg\,{\sc ii} absorption strength on scales within $1.4$ arcsec of the separation between the four lines of sight, which corresponds to the transverse proper size of about $12$ kpc.

In the spectra, the Mg\,{\sc ii} $\lambda \lambda 2796, 2803$ doublet absorption lines nearly fall at the wavelength corresponding to that of the H\,{\sc i} absorption line at $z_{\rm abs}=1.662$, which was previously identified in the HST-FOS observation \citep{Monier09}.  
In Table $\ref{table_EW}$, we also present the H\,{\sc i} column densities of the H\,{\sc i} absorption lines for the individual components 
in the four A/B/C/D spectra.  
\citet{Monier09} found that the H\,{\sc i} column densities typically change by factors of $2-20$: B $>$ A $>$ C $>$ D. 
The H\,{\sc i} absorption system at $z_{\rm abs}=1.662$ shown in the FOS spectrum toward component B is a typical DLA system with the highest  H\,{\sc i} column density, $N_{\rm HI} \sim 6.0 \times 10^{20}$ cm$^{-2}$, while the H\,{\sc i} column density in the spectrum toward  component D is the lowest, 
$N_{\rm HI}$ $\sim$ $0.30 \times 10^{20}$ cm$^{-2}$.
The variation in H\,{\sc i} column density is likely similar to that in Mg\,{\sc ii} absorption strength, which also exhibits the equivalent widths changing as B $>$ A $>$ C $>$ D. 
In Section 4, we will draw a detailed comparison of the variations between the lines of sight toward quasar H1413+1143.

For the Mg\,{\sc ii} absorption line No.1a, we obtain doublet ratios (DRs) of the equivalent width of the absorption line at $\lambda = 2796$ ${\rm \AA}$ to that at $\lambda = 2803$ ${\rm \AA}$, DR ($=W_{\rm MgII}^{0}(\lambda 2796)/W_{\rm MgII}^{0}(\lambda 2803)$) $=1.10\pm0.04, 1.20\pm0.05, 1.11\pm0.06$, and $1.05\pm0.06$ in the A, B, C, and D spectra, respectively. 
For the line No.1b,  DR$=1.03\pm0.10, 1.15\pm0.10, 0.91\pm0.12$, and $0.59\pm0.26$ in the A, B, C, and D spectra, respectively. 
In Section 4, we will discuss the implications from the DRs between the Mg\,{\sc ii} systems in the four lines of sight.

We also identify two possible Mg\,{\sc i} $\lambda 2853$ absorption lines at  $z_{\rm abs} = 1.659$ (No.2) and $1.667$ (No.3) in {\it all the four separate spectra.}  
The rest equivalent widths of Mg\,{\sc i} absorption lines are shown in Table $\ref{table_EW}$. 
The equivalent widths of line No.2 change by factors of $\la 2$: B $>$ A $\sim$ D $>$ C, while for line No.3, the equivalent widths change by factors of $\la 1.7$:  C $>$ D $\sim$ A $>$ B. 
It should be noted that line No.2 is properly aligned with the corresponding Mg\,{\sc ii} system at $z_{\rm abs} = 1.660$.  
However, line No.3 could be due to a different transition arising from a system at a different redshift,  
since it has a relatively large velocity difference 
($\ga 300$ km s$^{-1}$) from the Mg\,{\sc ii}  system at $z_{\rm abs} = 1.664$. 
The high-resolution spectra will be required to determine the origin of the velocity difference more correctly.

\subsection{Mg\,{\sc ii} System at $z_{\rm abs}=2.069$}

In Figure $\ref{spec_figure_new}$ (right panels), we find strong Mg\,{\sc ii} absorption doublets (No.4 $\&$ 5) at $z_{\rm abs} = 2.069$ in the A/B/C spectra. 
By contrast, no Mg\,{\sc ii} absorption line is detected in the D spectrum. 
The rest equivalent width of the Mg\,{\sc ii} line in the A spectrum,  
$W_{\rm MgII}^{0}(\lambda 2796)$ $=0.54 \pm 0.01$ ${\rm \AA}$, is the largest in the three absorption lines. 
The rest equivalent widths change by factors of up to $\sim 2$; A $>$ C $>$ B. 
Similar to the Mg\,{\sc ii} systems at $z_{\rm abs} \sim 1.66$, a presence of the variation in Mg\,{\sc ii} absorption strength is found for the  system at $z_{\rm abs}=2.069$. 
In the A, B, and C spectra,  the Mg\,{\sc ii} doublets exhibit the ratios DRs of $1.35\pm0.06$, $1.72\pm0.18$, and $2.00\pm0.33$, respectively. 
In Section 4, we will make a detailed comparison of the variations in absorption strength between the lines of sight and address the difference in DR of the Mg\,{\sc ii} doublets.    

\subsection{Mg\,{\sc ii} System at $z_{\rm abs}=2.097$}

In Figure $\ref{spec_figure_new}$ (right panels), we also find a Mg\,{\sc ii} absorption doublet (No.6 $\&$ 7) at  $z_{\rm abs} = 2.097$ in the  B spectrum, while no Mg\,{\sc ii} absorption line is detected in the other A/C/D spectra. 
The Mg\,{\sc ii} absorption line presents the rest equivalent width of $W_{\rm MgII}^{0}(\lambda 2796)$ $=$ $0.20 \pm 0.01$ ${\rm \AA}$ in the B spectrum. 
It should be noted that this system at $z_{\rm abs}=2.097$ is also identified as a sub-DLA system with H\,{\sc i} column density  $N_{\rm HI} \sim 5 \times 10^{19}$ cm$^{-2}$ which rise a strong Ly$\alpha$ absorption line with the rest equivalent width of $2.89 \pm 0.06$ ${\rm \AA}$ in the B spectrum \citep{Monier98}. 
In the other A, C, and D spectra, the strong H\,{\sc i} absorptions are also identified with the rest equivalent widths 
$W_{\rm HI}^{\rm 0}$ $=$ $1.32\pm0.08$  ${\rm \AA}$, $1.19 \pm 0.08$ ${\rm \AA}$, and $1.34\pm0.07$ ${\rm \AA}$, respectively. 
The sub-DLA system shows little difference in the H\,{\sc i} column densities between the three A/C/D spectra.  
This suggests that the system at $z_{\rm abs}=2.097$ in the B spectrum shows the strongest H\,{\sc i} absorptions in the four spectra, which is consistent with our result that the Mg\,{\sc ii} absorption line  is identified only in the B spectrum. 
For this system in the B spectrum, we obtain a doublet ratio DR of $1.82\pm0.26$. 
In Section 4, we further investigate the differences in absorption strength of the lines including the Mg\,{\sc ii} absorption line.  

%%%%%%%%%%%%%%%%%%%%%%%%%

\section{Discussion}

The multiple and coincident absorption lines offer valuable information about the chemical and dynamical structures of the intervening system along the sightlines toward background radiation sources. 
In particular, the quadruply gravitationally lensed quasar H1413+1143 provides the four separate spectra which produce several absorption lines on scales of ten kilo-parsecs between the lines of sight. 
In this section, we discuss the variation of Mg\,{\sc ii} absorption lines in the spatially resolved four spectra obtained by the Kyoto 3D\,{\sc ii} in comparison to those of the other absorption lines identified in the previous observations. 

\subsection{Mg\,{\sc ii} systems at $z_{\rm abs}=1.66$}

\begin{figure}[h]
\begin{center}
%\includegraphics[width=10cm]{/Users/okosh/Documents/data/z_abs_16595_AB_ver2.pdf}
%[two column]\includegraphics[width=10cm]{fig4.pdf}
\includegraphics[width=8.5cm]{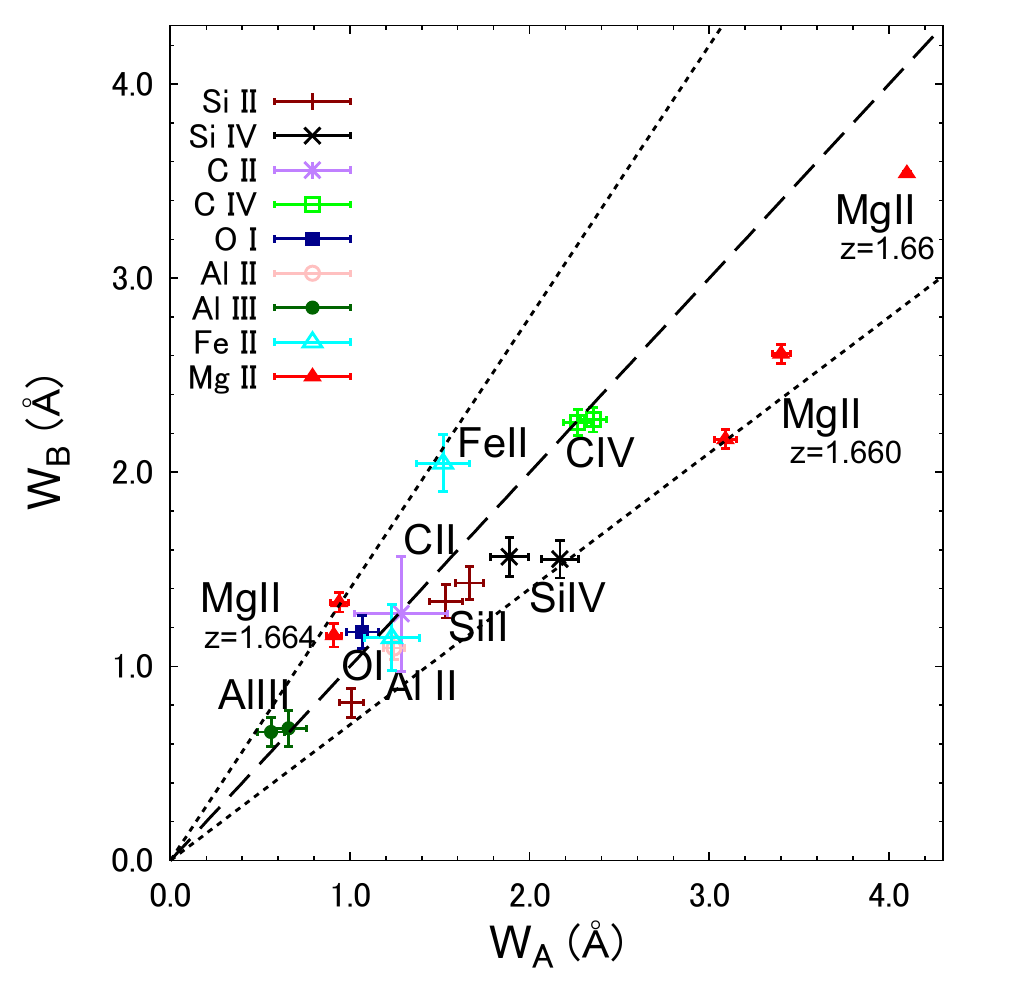}
\end{center}
\caption{Rest equivalent widths (EWs) of Mg\,{\sc ii} $\lambda \lambda 2796, 2803$ doublet absorption lines (red) at $z_{\rm abs}=1.660$ and $1.664$ in comparison to the other metal  absorption lines in the two A and B spectra. The data with $1 \sigma$ errors indicate the EW of metal absorption lines:   
Mg\,{\sc ii} (red), Si\,{\sc ii}(dark red), Si\,{\sc iv}(black), C\,{\sc ii}(purple), C\,{\sc iv}(green), O\,{\sc i}(dark blue), Al\,{\sc ii}(pink), Al\,{\sc iii}(dark green), and Fe\,{\sc ii}(cyan) \citep{Monier98}. The rest EW of Mg\,{\sc ii} absorption line at $z_{\rm abs}=1.66$, which is a half of the {\it total} EW, is also plotted (red). 
The relationship of the rest EWs between $EW(A)$ in the A spectrum and $EW(B)$ in the B spectrum; $EW(A) = EW(B)$ (dashed line) is shown together with $EW(B)=0.70EW(A)$(dotted  line) and $EW(B)=1.4EW(A)$(dotted  line).   
\label{z_abs_16595_AB_ver2}}
\end{figure}

In Figure $\ref{z_abs_16595_AB_ver2}$, we show the rest equivalent widths (EWs) of the Mg\,{\sc ii} $\lambda \lambda 2796, 2803$ absorption lines in the A and B spectra for comparing to those of the other metal  absorption lines at $z_{\rm abs} = 1.66$. 
We plot the rest EWs of a Mg\,{\sc ii} absorption line (No.1) at $z_{\rm abs}=1.66$ together with those of the Mg\,{\sc ii} $\lambda \lambda 2796, 2803$ doublets (No.1a $\&$ 1b) for the two components (red). 
For the rest EW of line No.1, we plot a half of the {\it total} EW ($=(W_{\rm MgII}^{0} (\lambda 2796)+W_{\rm MgII}^{0} (\lambda 2803))/2$) due to the blending of the doublet. 
Focusing on our result for the Mg\,{\sc ii} absorption lines (No.1, No1a, and No.1b), we find that the variation of the rest EWs in the  A and B spectra changes within factors between $0.7 \sim 1.4$ (dotted lines). 
For example, the No.1a line at  $z_{\rm abs} =1.660$ exhibits the ratio of EWs between the A and B spectra, $EW(B)/EW(A)$, of $\sim 0.7$. 
The No.1b line at $z_{\rm abs} = 1.664$ gives the ratio of $\sim 1.4$.

% Fig 4

The spectra in the previous observations provided various metal absorption lines at $z_{\rm abs}=1.66$: O\,{\sc i} $\lambda 1302$, C\,{\sc ii} $\lambda 1334$, Si\,{\sc iv} $\lambda \lambda 1393, 1402$, Si\,{\sc ii} $\lambda1526$,  C\,{\sc iv} $\lambda \lambda 1548, 1550$, Al\,{\sc ii} $\lambda 1670$, Al\,{\sc iii} $\lambda \lambda 1854, 1862$, and Fe\,{\sc ii} transitions \citep[e.g.][]{Hazard84, Drew84}. 
In the spatially resolved spectra toward four components of the H1413+1143 images, 
\citet{Monier98} investigated the difference in metal absorption strength between the sightlines using the HST-FOS.  
The observation allowed the precise measurement of EWs of the metal absorptions ({\it but Mg\,{\sc ii} doublets}). 
For example, in the A and B spectra, there is little difference between the EWs of the metal absorption lines.    
In Figure $\ref{z_abs_16595_AB_ver2}$, we also show the EWs of the metal absorption lines in the A and B spectra.
The result indicates that the EWs of low (e.g., C\,{\sc ii}, Si\,{\sc ii}) and high (e.g., C\,{\sc iv}, Si\,{\sc iv}) ions do not significantly differ from each other between the lines of sight to the images A/B. 
The EWs of the metal absorptions, including our measurements of the Mg\,{\sc ii} EWs, change by factors between $0.7$ and $1.4$. 
We find that the degree of the variation of Mg\,{\sc ii} EWs between the A and B spectra is not significantly different from those of the other metal EWs.

% Fig 5
%[two column]\begin{figure}[h]
\begin{figure*}[t]
\begin{center}
\includegraphics[width=18cm]{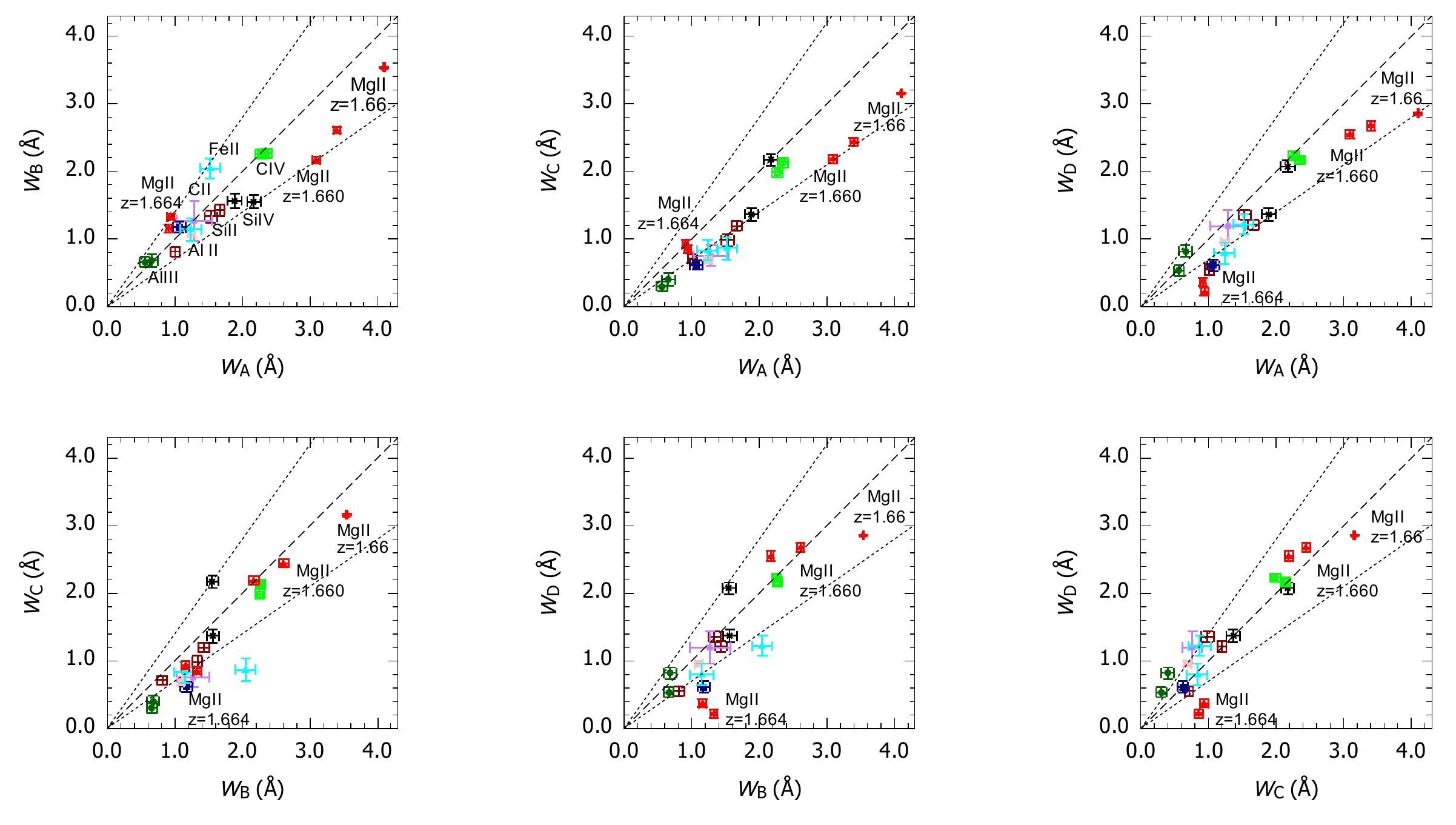}
\end{center}
\caption{Rest equivalent widths of Mg\,{\sc ii} doublet absorption lines (red) at $z_{\rm abs}=1.660$ and $1.664$ in comparison to the other metal  absorption lines in the spectra for the combinations of the four separate sightlines in the A/B/C/D spectra. 
The data with $1 \sigma$ errors indicate the EWs of the metal absorption lines:   
Mg\,{\sc ii} (red), Si\,{\sc ii}(dark red), Si\,{\sc iv}(black), C\,{\sc ii}(purple), C\,{\sc iv}(green), O\,{\sc i}(dark blue), Al\,{\sc ii}(pink), Al\,{\sc iii}(dark green), and Fe\,{\sc ii}(cyan) \citep{Monier98}.
The EW of Mg\,{\sc ii} absorption line at $z_{\rm abs}=1.66$, which is a half of the {\it total} EW, is also shown (red).
The relationship of the rest EWs between $EW(A)$ in the A spectrum and $EW(B)$ in the B spectrum: $EW(A) = EW(B)$ (dashed line) is shown together with $EW(B)=0.70EW(A)$(dotted  line) and $EW(B)=1.4EW(A)$(dotted  line).    
\label{z_abs_16595_ver2}}
\end{figure*}

In Figure $\ref{z_abs_16595_ver2}$, we summarize the EWs of various metal absorption lines, including our measurements of Mg\,{\sc ii} absorption lines (red),  in the pair-lines of sight toward four images A/B/C/D. 
For the metal absorptions but Mg\,{\sc ii}, \citet{Monier98} noted that the A spectrum shows the strongest absorption lines in the four spectra.  
Our result also shows that the strongest Mg\,{\sc ii} absorption line (No.1) at $z_{\rm abs}=1.66$ is detected in the A spectrum (Table $\ref{table_EW}$).  
Similar to the results for the variation of the EWs between the A and B spectra in Figure $\ref{z_abs_16595_AB_ver2}$, the EW variations of metal absorptions do not significantly differ from that of the Mg\,{\sc ii} absorption line (No.1) at $z_{\rm abs}=1.66$ (red) between the A/B/C/D spectra. 
Furthermore, the rest EWs of the Mg\,{\sc ii} $\lambda \lambda 2796, 2803$ doublet in line No.1a at $z_{\rm abs}=1.660$ (red) are also plotted. 
We find the variation in EW of the Mg\,{\sc ii} absorption line (No.1a) is similar to that of the blended line (No.1). 
For example, the absorption line No.1a in the A spectrum also shows the largest EW (Table $\ref{table_EW}$). 
This indicates that Mg\,{\sc ii} absorption line (No.1a) at $z_{\rm abs}=1.660$ likely arises from the absorber, which also gives rise to the other metal absorption lines previously identified at $z_{\rm abs}=1.660$ \citep{Monier98}.  
By contrast, for the No.1b line at $z_{\rm abs}=1.664$, the B spectrum shows the strongest absorption, while the D spectrum shows the weakest, as mentioned in the previous section. 
In Figure $\ref{z_abs_16595_ver2}$, the EWs of the Mg\,{\sc ii} doublets at $z_{\rm abs}=1.664$ (red) are also shown, together with those of the other metal lines at $z_{\rm abs}=1.660$. 
We find that, for line No.1b, the EW differences between the lines of sight to the images A/B/C are similar to those of metal lines at $z_{\rm abs}=1.660$. 
However, the difference between the D spectrum and the other A/B/C spectra appears to be large due to the small EWs of the Mg\,{\sc ii} doublets No.1b at $z_{\rm abs}=1.664$ in the D spectrum.

\begin{figure}[h]
\begin{center}
%\includegraphics[width=10cm]{/Users/okosh/Documents/data/ratio_z16595_ver4.pdf}
%[two column]\includegraphics[width=10cm]{fig6.pdf}
\includegraphics[width=8.5cm]{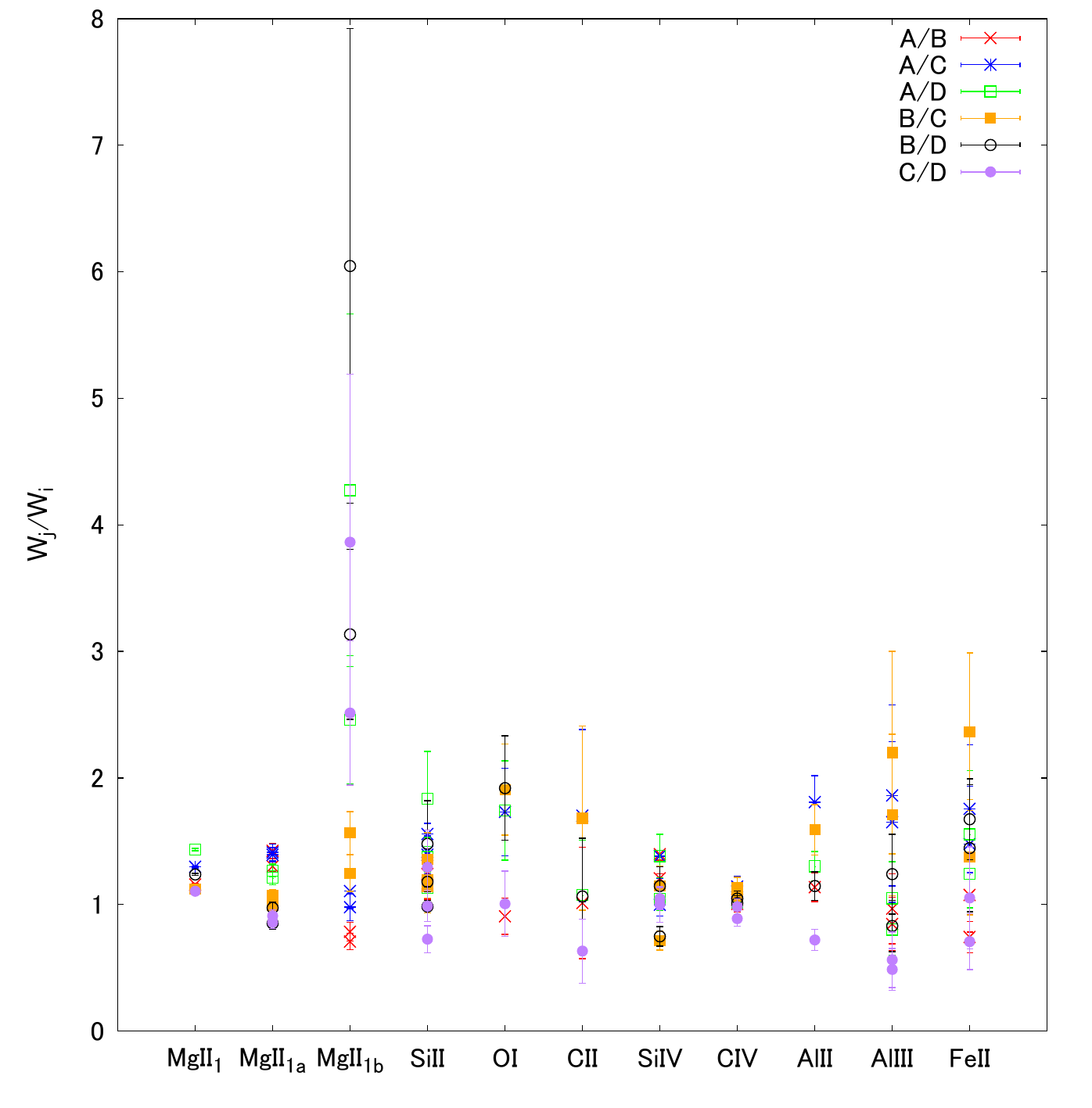}
\end{center}
\caption{Ratios between the equivalent widths of the metal absorption lines at $z_{\rm abs}  \sim 1.66$ in the four A/B/C/D  spectra together with those of the other metal absorption lines. The ratios of the EW in the A spectrum to the  EW in the B spectrum are denoted as points (red). The error bars indicate $1 \sigma$ errors of the EW of each absorption line. 
The EW ratios A/C, A/D, B/C, B/D, and C/D are also shown as points (blue, green, orange, black, and purple,  respectively).   
The ratios between the EWs of the Mg\,{\sc ii} line (No.1) at $z_{\rm abs}=1.66$  are shown as 
`Mg\,{\sc ii}$_{1}$'. 
Also shown are the ratios between the EWs of the Mg\,{\sc ii} doublets (No.1a $\&$ 1b) at $z_{\rm abs}=1.660$ and $1.664$ as `Mg\,{\sc ii}$_{1a}$' and `Mg\,{\sc ii}$_{1b}$', respectively. 
\label{ratio_z16595_ver4}}
\end{figure}

Here we also focus on a degree of the variation in EW between the two spectra in the four lines of sight toward components A/B/C/D more precisely. 
In Figure $\ref{ratio_z16595_ver4}$, we show the ratios between the Mg\,{\sc ii} EWs in the pair-sightlines toward four images A/B/C/D, together with those of the other metal absorption lines. 
Our measurements of Mg\,{\sc ii} absorption No.1, No.1a, and No.1b are denoted as Mg\,{\sc ii}$_{\rm 1}$, Mg\,{\sc ii}$_{\rm 1a}$, and Mg\,{\sc ii}$_{\rm 1b}$, respectively. 
The ratios between the EWs of the Mg\,{\sc ii} absorption lines (No.1) at $z_{\rm abs}=1.66$ in the pair of lines of sight toward the four images A/B/C/D change by factors between $1.0$ and $1.4$, while those of the other metal absorption lines change by factors between $0.5$ and $2.4$. 

% Fig 6

For the No.1b line at $z_{\rm abs}=1.664$,    
we find that the EW ratios change by factors of up to $6$, which obviously differs from those of Mg\,{\sc ii} and the other metal absorption lines at $z_{\rm abs} =1.660$. 
The Mg\,{\sc ii}-EW in the D spectrum is obviously smaller than those in the A/B/C spectra.  This mainly causes the large ratios of Mg\,{\sc ii}-EWs shown in Figure $\ref{ratio_z16595_ver4}$.   
The large ratios suggest that the physical properties of component No.1b is not likely similar to those of component No.1a.   
The No.1b component at $z_{\rm abs}=1.664$ may not be associated with the absorbers at $z_{\rm abs}=1.660$   
since the comoving distance between the two systems is $\sim 7$ Mpc along the line of sight assuming only the Hubble expansion.

In the four A/B/C/D spectra, the column densities of H\,{\sc i} absorption lines change by factors of up to $20$ between the B and D spectra  (Table $\ref{table_EW}$). 
The high degree of the variation in H\,{\sc i} absorption strength is similar to  those of the Mg\,{\sc ii} absorptions (No.1b) at $z_{\rm abs} =1.664$. 
Furthermore, for the system at $z_{\rm abs} =1.664$, the EWs of Mg\,{\sc ii} absorption change as B $>$ A $>$ C $>$ D, which is qualitatively equivalent to the decreasing order in the H\,{\sc i} absorption strength (Table $\ref{table_EW}$).  
Although it is still unclear that the DLA system arises which Mg\,{\sc ii} absorptions at $z_{\rm abs} =1.660$ or $1.664$, 
the results indicate {\it an inhomogeneous spatial distribution of Mg\,{\sc ii} and H\,{\sc i} absorbers} at $z \sim 1.66$ on scales of the separation in the transverse direction between the multiple background images A/B/C/D within $1.4$ arcsec corresponding to the physical scale of $\sim 12$ kpc. 
The spatial distribution of the multiple and coincident absorptions indicates that Mg\,{\sc ii} absorbers trace inhomogeneous H\,{\sc i} absorbers (DLA systems) on scales within $\sim 12$ kpc in the CGM 
if a galaxy is embedded in the halos with CGM, clouds, or galactic flows. 
For example, recent studies of the CGM have focused on the galactic feedback process through a relationship between H\,{\sc i} absorbers and metal ones.  
They often provided strong evidence that low-ionization metal absorption systems, in particular, strong  Mg\,{\sc ii} systems with large equivalent width $W_{\rm MgII}^{0} > 1$ ${\rm \AA}$, originate from galactic outflows  \citep[e.g.][]{Nestor11, Bouche12, Turner14}. 
Moreover, the Mg\,{\sc ii}-selected DLAs at $0.4 < z_{\rm abs} < 1.5$ show a correlation between the H\,{\sc i} column density and the rest EWs of Mg\,{\sc ii}, $ 0.5 < W_{\rm MgII}^{0}(\lambda 2796)/{\rm \AA} < 3$ \citep{Menard09} . 
Specifically, the high-$N_{\rm HI}$ systems likely exhibit strong Mg\,{\sc ii} absorptions. 
This is consistent with the variations in H\,{\sc i} and Mg\,{\sc ii} absorption strengths in the spectra of quasar H1413+1143.  
In the four lines of sight to the H1413+1143 images, the inhomogeneous metal-enrich H\,{\sc i} gas gives rise to the DLA absorption lines, together with the Mg\,{\sc ii} absorptions.

\subsection{Mg\,{\sc ii} system at $z_{\rm abs}=2.069$}

\begin{figure*}[t]
\begin{center}
\includegraphics[width=18cm]{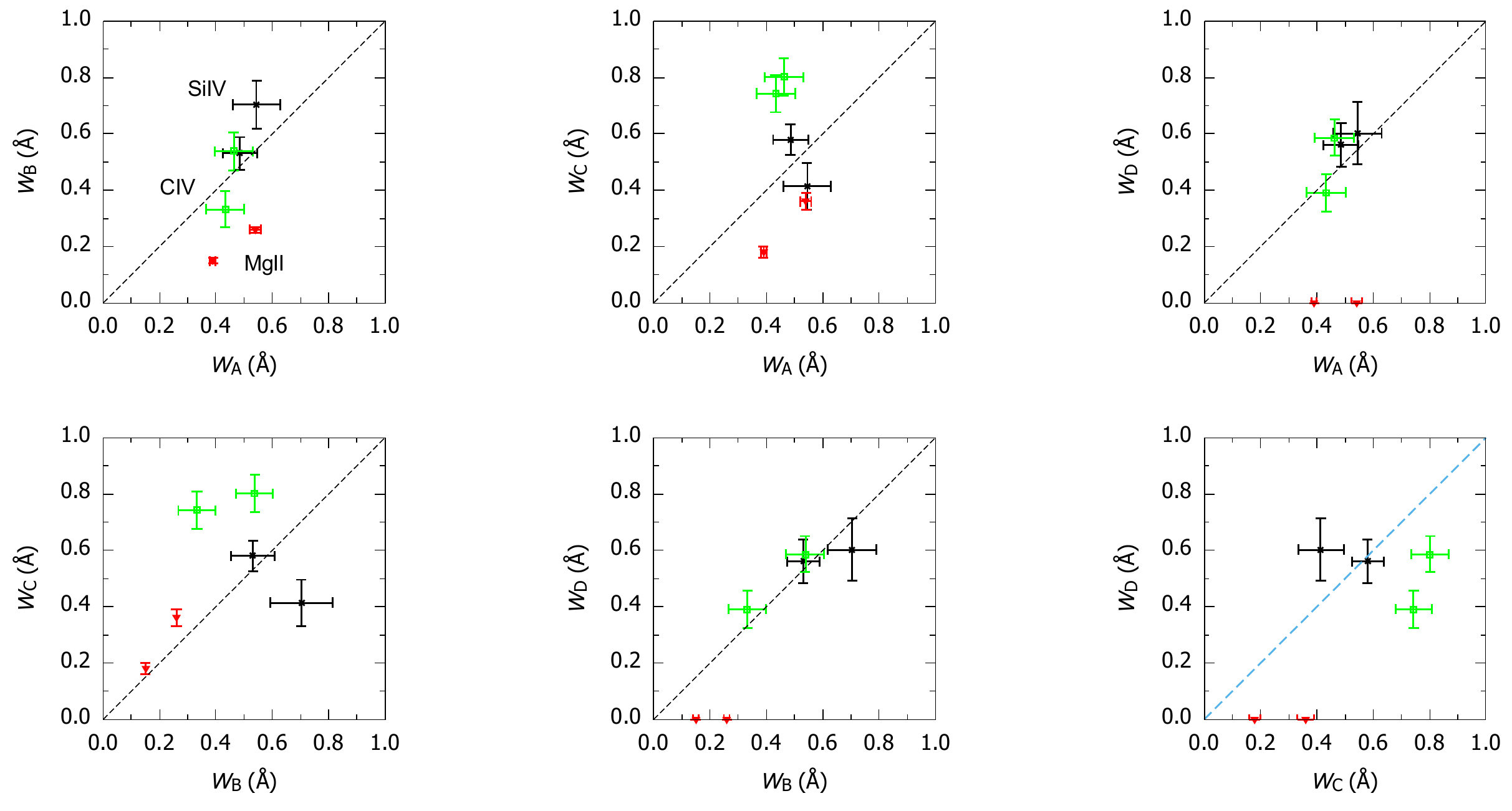}
\end{center}
\caption{Rest equivalent widths of the Mg\,{\sc ii} doublet absorption lines (red) at $z_{\rm abs}=2.069$ in comparison to the other metal  absorption systems, Si\,{\sc iv}(black) and C\,{\sc iv}(green) \citep{Monier98}, in the spectra for the combinations of the four separate sightlines in the A/B/C/D spectra. 
The data with $1 \sigma$ errors indicate the EWs of metal absorption lines.
The relationship of the rest EWs between $EW(i)$ in the spectrum $i$ and $EW(j)$ in the spectrum $j$, $EW(i) = EW(j)$,  is shown as a dashed line. 
\label{z_abs_20680}}
\end{figure*}

In Figure $\ref{z_abs_20680}$, 
% Fig 7
%[two column]\begin{figure}[h]
we show the EWs of metal absorption lines including our measurement of the low-ion doublet Mg\,{\sc ii} $\lambda \lambda 2796, 2803$ (No.4 $\&$ 5) at $z_{\rm abs}=2.069$ (red) in the pair of sightlines toward four images A/B/C/D. 
\citet{Monier98} identified the high-ions doublets Si\,{\sc iv} $\lambda \lambda 1393, 1402$ (black) and C\,{\sc iv} $\lambda \lambda 1548, 1550$ (green) at $z_{\rm abs}=2.068$ 
and made comparisons of the rest EWs between the four lines of sight. 
For the C\,{\sc iv} doublet, the absorption line is strongest in the C spectrum and approximately the same strength in the A, B, and D spectra. 
Our result reveals that the absorption strengths of the low-ion Mg\,{\sc ii} doublet lines changes as A $>$ C $>$ B, with no detection in D. 
This clearly differs from the variations of the high-ion absorption strength.

\begin{figure}[h]
\begin{center}
%\includegraphics[width=10cm]{/Users/okosh/Documents/data/ratio_z20680_ver2.pdf}
%[two column]\includegraphics[width=10cm]{fig8.pdf}
\includegraphics[width=8.5cm]{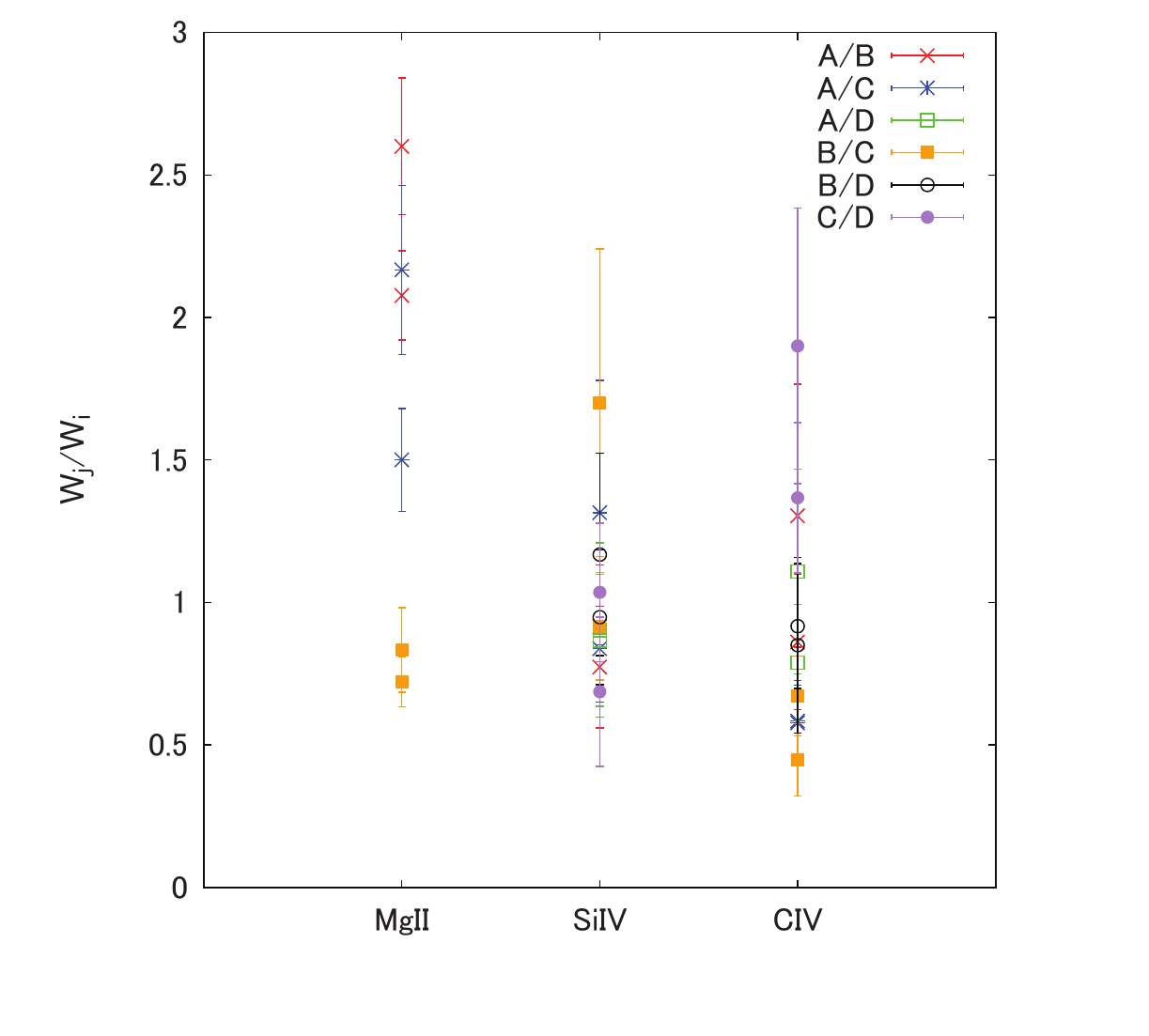}
\end{center}
\caption{Ratios between the EWs of metal absorption lines at $z_{\rm abs} =2.069$ in the four A/B/C/D spectra,  together with those of the other metal absorption lines. The ratios of the EW in the A spectrum to the EW in the B spectrum are denoted as points (red).  The error bars indicate $1 \sigma$ errors of the EW of each absorption line. 
The EW ratios A/C, A/D, B/C, B/D, and C/D are also shown as points (blue, green, orange, black, and purple, respectively) with $1 \sigma$ errors.  
\label{ratio_z20680_ver2}}
\end{figure}

For a detailed comparison of the EWs between the four lines of sight, in Figure $\ref{ratio_z20680_ver2}$, we show the ratios of the EWs for the Mg\,{\sc ii}, C\,{\sc iv}, and Si\,{\sc iv} doublets in the pair-sightlines. 
We find that the ratios of the Mg\,{\sc ii} EWs in the pair-lines of sight A/B/C change by factors between $0.7$ and $2.6$, which do not significantly differ from those of the high-ion doublets of C\,{\sc iv} and Si\,{\sc iv} between $0.5$ and $2.0$. 
However, in the D spectrum, the Mg\,{\sc ii} doublet has not been detected, whereas the C\,{\sc iv} and Si\,{\sc iv} doublets are detected with rest EWs of $\sim 0.4-0.6 {\rm \AA}$, which might be due to partly blending with the other lines associated with quasar H1413+1143 \citep{Monier98}. 
The difference in the Mg\,{\sc ii} EW indicates that, in the system at $z_{\rm abs}=2.069$, either (1) the cross section and/or filling factor of the absorbers is small, or   
(2) the absorber in the line of sight toward component D is highly ionized in the radiation field, which is different from that ionizing the absorbers in lines of sight toward components A/B/C.
Due to the small number statistics, it is still not clear what causes the variation of the EWs between the four lines of sight. 
Further investigation for the system at $z_{\rm abs}=2.069$ requires more samples of the other high-/low-ion absorption lines 
to put constraints on the ionization structure.

% Fig 8

\subsection{Mg\,{\sc ii} systems at $z_{\rm abs}=2.097$}

\begin{figure*}[t]
\begin{center}
\includegraphics[width=18cm]{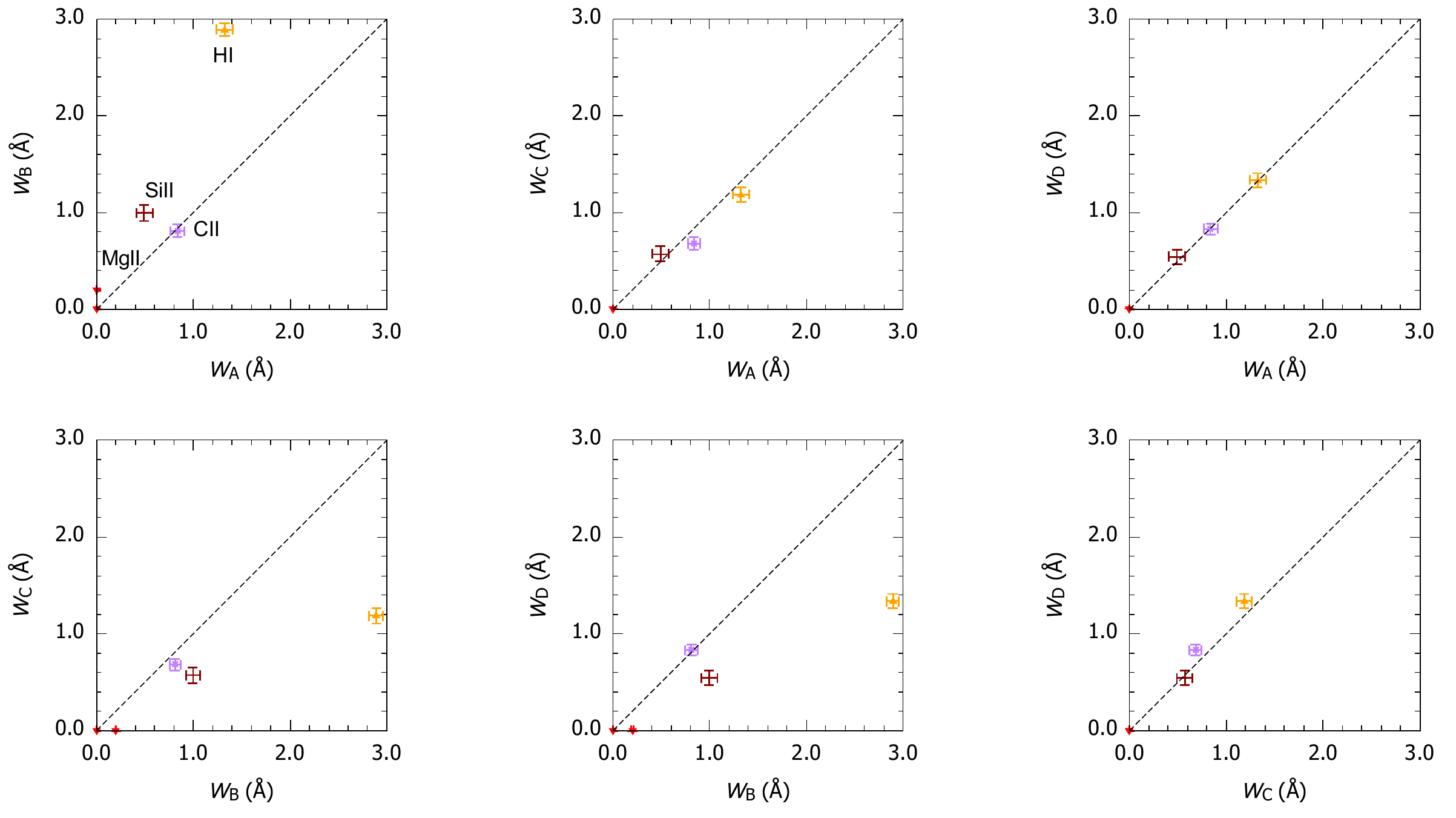}
\end{center}
\caption{Rest equivalent widths of Mg\,{\sc ii} doublet absorption lines (red) at $z_{\rm abs}=2.097$ in comparison to the other metal absorption systems, Si\,{\sc ii}(dark red), C\,{\sc ii}(purple), and H\,{\sc i}(orange) \citep{Monier98}, in the spectra for the combinations of the four separate A/B/C/D spectra.  
The data with $1 \sigma$ errors indicate the EWs of metal absorption lines.  
The relationship of the rest EWs between $EW(i)$ in the spectrum $i$ and $EW(j)$ in the spectrum $j$: $EW(i) = EW(j)$  is shown as a dashed line. 
\label{z_abs_20969}}
\end{figure*}

In Figure $\ref{z_abs_20969}$, we show the EWs of metal absorption lines, including our measurement of the Mg\,{\sc ii} doublet (No.6 $\&$ 7) at $z_{\rm abs}=2.097$, in the pair-sightlines toward four images A/B/C/D. 
The Mg\,{\sc ii} doublet (No.6 $\&$ 7) at  $z_{\rm abs} = 2.097$ is identified in the B spectrum while {\it no} Mg\,{\sc ii} absorption line is detected in the other A/C/D spectra (Table $\ref{table_EW}$).  
The rest EW of the Mg\,{\sc ii} doublet is plotted (red).
This system also gives rise to the low-ion absorption lines, Si\,{\sc ii} $\lambda 1260$ and C\,{\sc ii} $\lambda 1334$, in the four lines of sight. 
Similar to the variation in H\,{\sc i} absorption strength in the four sightlines that is addressed in Section 3, the Si\,{\sc ii} line is the strongest in the B spectrum, $EW_{\rm B} = 0.99 \pm 0.08 {\rm \AA}$,  
while $EW_{\rm A} = 0.49 \pm 0.08 {\rm \AA}$, $EW_{\rm C} = 0.47 \pm 0.08 {\rm \AA}$, and $EW_{\rm D} = 0.55 \pm 0.07 {\rm \AA}$ in the A, C, and D spectra, respectively \citep{Monier98}. 
The line of sight to the image B intersects the absorber giving rise to the strongest H\,{\sc i} and Si\,{\sc ii} absorption lines. 
This is consistent with our result for the Mg\,{\sc ii} absorption line clearly detected in the B spectrum but not in the other spectra. 
The low-ion Mg\,{\sc ii}/Si\,{\sc ii} absorption lines in the B spectrum are likely associated with the 
sub-DLA system at $z_{\rm abs}=2.097$.  
Similar to the system at $z_{\rm abs} = 2.069$, more samples of the other metal absorption lines (e.g., high-ion) are required to 
investigate the ionization structure of the sub-DLA system at $z_{\rm abs} = 2.097$.

% Fig 9
%[two column]\begin{figure}[htb]

\subsection{Multiple Mg\,{\sc ii} Absorption in Lines of Sight Toward Quadruply and Triply Lensed Quasars}

Multiple lines of sight toward background quasars offer a unique and valuable probe of the spatial structure of the intervening absorbers by mapping the transverse dimension. 
In particular, the gravitational lensing produces the quadruply or triply lensed images that provide the multiple lines of sight. 
Utilizing the advantage of the multiple sightlines given by the quadruply or triply lensed images, the scale of variations in absorption strength has been investigated in the previous studies \citep[e.g.][]{Rauch02, Ellison04, Oguri04, Chen14, Rubin18b}. 
In Table $\ref{table_QSO}$, we present a compilation of the multiple Mg\,{\sc ii} absorption EWs at $\lambda$ $=2796$ ${\rm \AA}$ in lines of sight toward quadruply and triply lensed quasars that we have used for comparing to our measurements here.
The transverse distances $d_{\rm max}$ and $d_{\rm min}$ present the largest and smallest proper separations at the absorption redshifts between the lines of sight toward the images, respectively. 
For quasar H1413+1143, $d_{\rm max}$ is the largest separation in the transverse direction between the line of sight to image B and that to image C, and $d_{\rm min}$ is the smallest one between the lines of sight to images A and B.    

% Fig 10
\begin{figure}[h]
\begin{center}
\includegraphics[width=8.5cm]{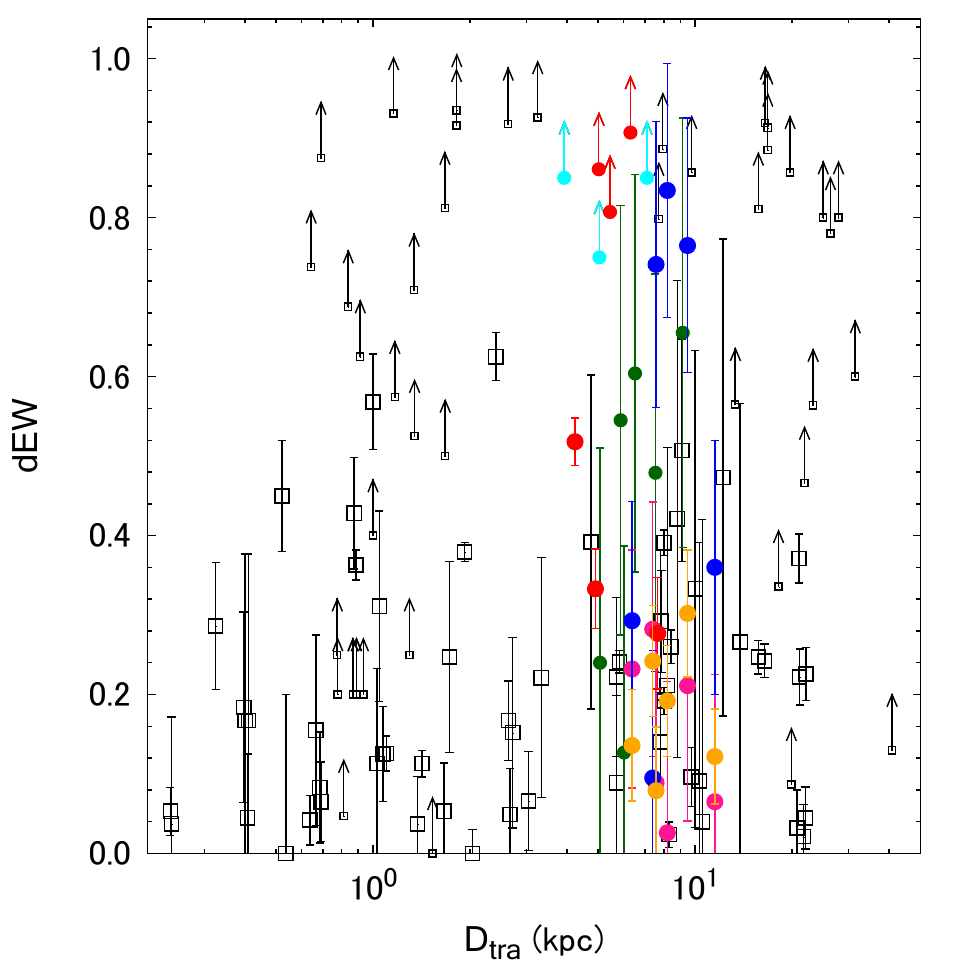}
\end{center}
\caption{Fractional equivalent width difference $dEW$ as a function of physical separation in the transverse direction $D_{\rm tra}$ for the Mg\,{\sc ii} absorption systems in the H1413+1143 spectra at 
$z_{\rm abs}=1.66$ (orange) (the two components $z_{\rm abs}=1.660$ (pink) $\&$ $1.664$ (blue)), $2.069$ (red), $2.097$ (cyan), and $0.609$ (green) together with  the other systems (black open-squares).  
\label{dEW_D_sample1}}
\end{figure}

\subsubsection{Fractional difference $dEW$ versus Transverse separation $D_{\rm tra}$}

In Figure $\ref{dEW_D_sample1}$,  we present the fractional difference, $dEW$, in Mg\,{\sc ii} EW for each pair of the sightlines as a function of physical separation in the transverse direction $D_{\rm tra}$, 
\begin{eqnarray}
dEW & = \frac{\displaystyle W_{\rm r}^{\rm X} -  W_{\rm r}^{\rm Y}}{\displaystyle W_{\rm r}^{\rm X}}, \nonumber
\end{eqnarray}
where $W_{\rm r}^{\rm X}$ and $W_{\rm r}^{\rm Y}$ are the rest EWs at $\lambda=2796$ ${\rm \AA}$ in the lines of sight toward the separate images $X$ and $Y$, respectively, 
in a case that $W_{\rm r}^{\rm X} > W_{\rm r}^{\rm Y}$. 
For the quadruply lensed quasar, the four sight lines give six pairs of sightlines that provide six measurements of $dEW$. 
The left panel of Figure $\ref{dEW_D_sample1}$ shows our measurements of $dEW$  (filled circles) for the Mg\,{\sc ii} absorption lines in the spectra of the quadruply lensed quasar H1413+1143 at $z_{\rm abs}=1.66$ (No.1) (orange), $2.069$ (red), and $2.097$ (cyan), together with the previous measurement of $dEW$ at $z_{\rm abs}=0.609$ (green) \citep{Monier98}. 
For the Mg\,{\sc ii} systems at $z_{\rm abs}=1.66$ (orange), we find that the fractional differences $dEW$ are small, $\sim 30 \%$ or less ($0.07 < dEW <0.30$) with the transverse separations between the lines of sight of $6$ kpc $< D_{\rm tra}$ $<$ $12$ kpc. 
For the systems that include no detection yielding lower limits at $z_{\rm abs}=2.069$ (red) and $2.097$ (cyan),   $dEW > 0.27$ and $dEW > 0.75$, respectively.  
The system at $z_{\rm abs}=0.609$ (green) exhibits $0.12 < dEW <0.65$.
For the two components of line No.1 at $z_{\rm abs}=1.66$, we also plot the measurements of $dEW$ of the lines 
No.1a (pink) and No.1b (blue):  
$0.02 < dEW <0.28$ for the system at $z_{\rm abs}=1.660$ (pink) and $0.09 < dEW <0.83$  for one at $z_{\rm abs}=1.664$ (blue). 
For component No.1a at $z_{\rm abs}=1.660$ (pink), the fractional differences $dEW$ are small, $\sim 30 \%$ or less, 
which are similar to those of the blended line No.1. 
By contrast, the fractional differences $dEW$ for component No.1b at $z_{\rm abs}=1.664$ (blue) have large scatters,  
which are previously shown for the ratios of EWs in the pair lines of sight in Figure $\ref{ratio_z16595_ver4}$.

In Figure $\ref{dEW_D_sample1}$,  we include the previous measurements for the other pairs of  Mg\,{\sc ii} absorption systems in the spectra of the doubly, triply, and quadruply lensed quasars and pair quasars (black open squares) in the literature 
\footnote[1]{
The studies used different cosmological parameters in order to estimate the physical separations of the quasar images for each absorption system. Several studies adopted a cosmological model with 
the deceleration parameter $q_{\rm 0}$ $=$ $0.5$, $\Omega_{\rm \Lambda} = 0$ and $H_{\rm 0} = 50$ km s$^{-1}$ Mpc$^{-1}$\citep[e.g.,][]{Crotts94, Lopez00, Rauch02}. 
For these samples, we calculate the physical separations of the quasar images by adopting the standard $\Lambda$CDM cosmological model ($\Omega_{\rm 0} = 0.3$, $\Omega_{\rm \Lambda} = 0.7$, and $H_{\rm 0} = 70$ km s$^{-1}$ Mpc$^{-1}$).
}
\citep[e.g.][]{Crotts94, Smette95, Lopez00, Rauch02, Churchill03, Ellison04, Oguri04, Oguri08, Chen14, Koyamada17, Rubin18b}.
The lower limits on $dEW$ indicate the fractional difference $dEW$ for systems ($\sim 40 \%$ of the total sample) with no detection in one line of sight. 
The total sample including our measurements, which include ones for lines No.1a $\&$ 1b instead of the composite line No.1 
has the transverse separations $D_{\rm tra}$ with a mean of  $15 \pm 2$ kpc and a median of $7.5$ kpc. 
We compare the $dEW$ distributions between two subsamples that have a transverse separation $D_{\rm tra}$ smaller than $7.5$ kpc and larger than $7.5$ kpc.  
To assess the significance of the differences in the $dEW$ distributions between the two subsamples, we perform a survival analysis. 
A log-rank test that compares the survival distributions between  the two subsamples yields a $p$-value of $\sim 0.0149$, 
ruling out the null hypothesis that the samples are drawn from the same parent population. 
The subsamples having the separations of $D_{\rm tra}$ $< 7.5$ kpc and $> 7.5$ kpc give the $dEW$ medians (survival) of $0.31$ and $0.80$, respectively.   
The absorbers with large $D_{\rm tra}$ likely have large fractional differences $dEWs$.
Similarly, \citet{Rubin18b} investigated the fractional differences $dEW$ for the multiple  
Mg\,{\sc ii} systems in the lensed quasar, including the quadruply lensed quasar J014710+463040. 
The authors found a trend that the fractional differences $dEW$ depend on the transverse distance $D_{\rm tra}$ (particularly $D_{\rm tra}$ $>10$ kpc)  
since the incident rate of $dEW$ smaller than $0.2$ decreases as the physical separation $D_{\rm tra}$ increases. 
Although the different detection limits in the literature make it difficult to compare the data with lower limits, the difference in the $dEW$ distributions may be attributed to the reason that nondetection systems yielding lower limits on $dEWs$ likely have relatively large $dEWs$. 
Indeed, the total sample including the lower limits gives a median (survival) of $dEW_{\rm median}$ $= 0.46$ which is twice as large as the one of $0.22$ for the subsample excluding the lower limits. 
The following are possible causes for the nondetection of absorption: (1) in a case where the absorber has a steep EW gradient, the large difference in EW between the separation between the lines of sight causes the no-detection of absorption in a line of sight due to the EW smaller than the detection limit, and (2) the line of sight does not penetrate the absorber, while the other line does. 
For example, the separations of the sightlines are much larger than the typical size of the absorbing system (e.g., single cloud or numerous small clouds), or the pairs of sightlines are not along with the absorber which structure is elongated or disklike. 
In this case, the samples with the lower limits could increase the values of $dEW$ (particularly for large $D_{\rm tra}$).  
This could cause the difference in the median of $dEW$ between the samples including the lower limits and those without the lower limits.

% Fig 11
\begin{figure}[h]
\begin{center}
%\includegraphics[width=10cm]{/Users/okosh/Documents/data/dEW_EW_sample1_only.pdf}
%[two column]\includegraphics[width=10cm]{fig11.pdf}
\includegraphics[width=8.5cm]{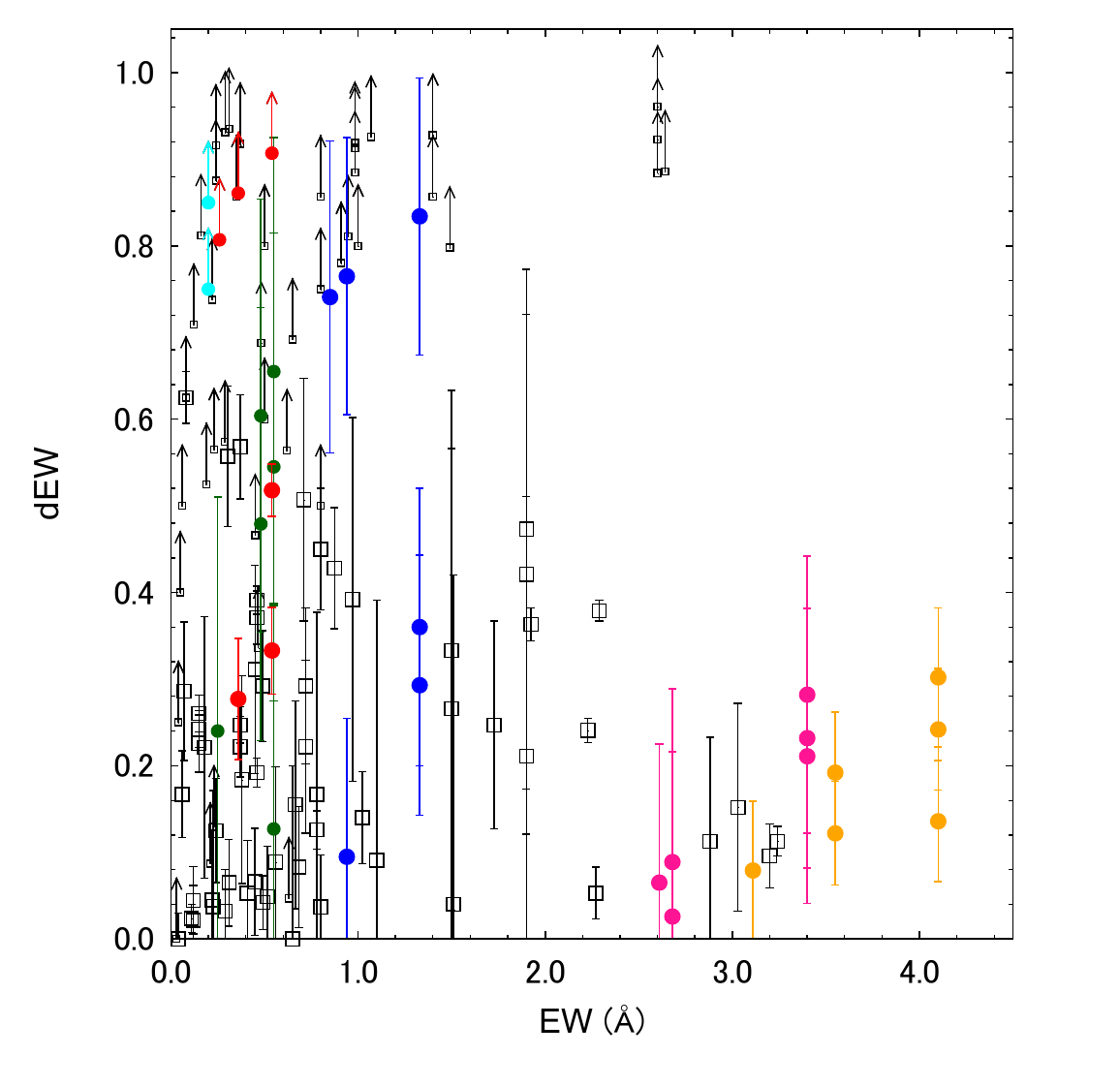}
\end{center}
%\caption{{\it Left}:
\caption{ Fractional equivalent width difference $dEW$ as a function of rest EW at $\lambda=2796$ ${\rm \AA}$ for the Mg\,{\sc ii} systems in the H1413+1143 spectra at $z_{\rm abs}=1.66$ (orange) (the two components $z_{\rm abs}=1.660$ (No.1a) (pink) $\&$ $1.664$ (No.1b) (blue)),  $2.069$ (red), $2.097$ (cyan), and 0.609 (green) together with  the other systems (black open squares). 
The equivalent width $EW$ is the larger one in the pair of Mg\,{\sc ii} absorption lines. 
\label{dEW_EW_sample1}}
\end{figure}

% Fig 12
%[two column]\begin{figure}[h]
\begin{figure*}[t]
\begin{center}
\includegraphics[width=15cm]{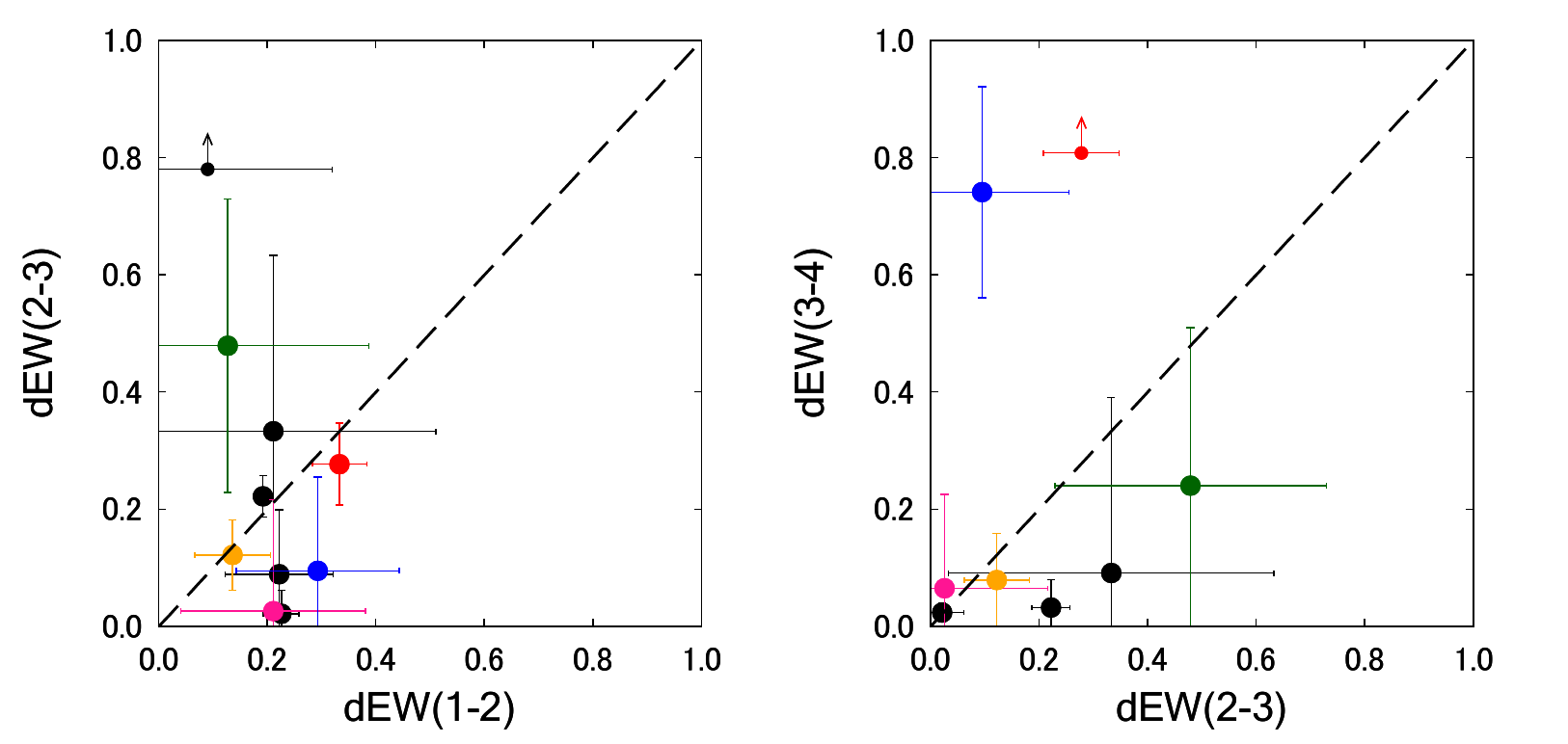}
\end{center}
\caption{
Left: Fractional equivalent width difference $dEW(1-2)$ and $dEW(2-3)$ for the absorption lines 
at $z_{\rm abs}=1.66$ (orange) (the two components $z_{\rm abs}=1.660$ (pink) and $1.664$ (blue)),  $2.069$ (red), and 0.609 (green) together with  the other systems (black). 
The systems which include no detection yielding lower limits are also shown.  
Here $dEW(1-2)$ is the fractional difference between the largest and 
second-largest EWs in the multiple lines of sight toward a quadruply lensed quasar, and    
$dEW(2-3)$ is the fractional difference between the second- and 
third-largest EWs in the multiple lines of sight toward a quadruply lensed quasar.   
The relationship of the rest EWs between $dEW(1-2)$ and $dEW(2-3)$, $dEW(1-2)$ $=$ $dEW(2-3)$ (dashed line),  is shown. 
Right: fractional equivalent width difference $dEW(2-3)$ and $dEW(3-4)$ for the absorption lines in the quadruply lensed quasars. 
The symbols are the same as the ones in the left panel.   
Here $dEW(3-4)$ is the fractional difference between the third- and 
fourth-largest (smallest) EWs in the four lines of sight toward a quadruply lensed quasar.   
\label{dEW_dEW}}
\end{figure*}

\subsubsection{Fractional Difference $dEW$ versus Equivalent Width $EW$}

Next, we focus on a relationship between the fractional difference $dEW$ and equivalent width EW of the Mg\,{\sc ii} absorption lines.   
In Figure $\ref{dEW_EW_sample1}$, we present the fractional difference $dEW$ as a function of the rest EW of Mg\,{\sc ii} absorption lines at $\lambda=2796$ ${\rm \AA}$ which is a larger one in the pair of Mg\,{\sc ii} absorption lines. 
Similar to Figure $\ref{dEW_D_sample1}$, we present our measurements of $dEW$ for the Mg\,{\sc ii} absorption lines in the spectra of the quadruply lensed quasar H1413+1143 at $z_{\rm abs}=1.66$ (No.1) (orange) 
(the two components $z_{\rm abs}=1.660$ (pink) $\&$ $1.664$ (blue)),  $2.069$ (red), and $2.097$ (cyan) together with the previous measurement of $dEW$ at $z_{\rm abs}=0.609$ (green) \citep{Monier98}. 
The result shows that the systems with small EWs ($< 2~{\rm \AA}$) have the fractional differences $dEW$ with large scatters.  
By contrast, the systems with large EWs have small scatters, which is likely due to the small sample size at $EW > 2$ ${\rm \AA}$. 
The total sample including our measurements, which include ones for lines No.1a $\&$ 1b instead of the composite line No.1, 
has the rest EWs with a mean of  $0.90 \pm 0.07$ ${\rm \AA}$ and a median of $0.55$ ${\rm \AA}$. 
We make two subsamples divided by the rest EW of $0.5$ ${\rm \AA}$.  
To assess the significance of the differences in the $dEW$ distributions between the two subsamples, we perform a survival analysis. 
A log-rank test comparing the survival distributions of the subsamples yields a $p$-value of $0.74$, 
which thus does not rule out the null hypothesis that the samples are drawn from the same parent population. 
Thus, we do not find evidence for a relationship between $dEW$ and $EW$, which is consistent with the result in \citet{Rubin18b}.

\subsubsection{Fractional Differences of Multiple Mg\,{\sc ii} absorption systems in the spectra of quadruply lensed quasars}

Here we focus on multiple Mg\,{\sc ii} absorption systems in the spectra of {\it only quadruply lensed quasars}. 
The spectra toward quadruply lensed quasars have the advantage of providing more than two measurements of $dEW$ 
in each single absorption system at the same redshift through three or more lines of sight toward the lensed quasars. 
For the quadruply lensed quasar H1413+1143, the four separate spectra provide six measurements of $dEW$ for the six pairs of 
the Mg\,{\sc ii} absorption lines at $z_{\rm abs}=1.66$ since {\it all} four lines of sight exhibit the Mg\,{\sc ii} absorption lines at $z_{\rm abs}=1.66$.  
In Figure $\ref{dEW_dEW}$, we compare the fractional differences $dEW$ {\it from the largest equivalent width} in descending order of EW in the three or four lines of sight toward the quadruply lensed quasars. 
For example, if a quadruply lensed quasar provides four EWs of the absorption lines in the four lines of sight,  
$W_{\rm 1}$, $W_{\rm 2}$, $W_{\rm 3}$, and $W_{\rm 4}$ ($W_{\rm 1} > W_{\rm 2} > W_{\rm 3} > W_{\rm 4}$), 
we compare the fractional differences $dEW(1-2)$ between $W_{\rm 1}$ and $W_{\rm 2}$ to $dEW(2-3)$ between $W_{\rm 2}$ and $W_{\rm 3}$. 
In the left panel, we show the relationship between $dEW(1-2)$ and $dEW(2-3)$ for the absorption lines 
in the three or four lines of sight toward the quadruply lensed quasars. 
In the right panel, we also present 
the fractional differences $dEW(2-3)$ between $W_{\rm 2}$ and $W_{\rm 3}$ in comparison to $dEW(3-4)$ between $W_{\rm 3}$ and $W_{\rm 4}$ in the four lines of sight toward quadruply lensed quasars. 
The median survival values of 
$dEW(1-2)$, $dEW(2-3)$, and $dEW(3-4)$ are estimated to be 
$0.21^{+0.01}_{-0.02}$, $0.17^{+0.11}_{-0.08}$, and $0.085^{+0.155}_{-0.020}$, respectively. 
Here the uncertainties denote 
the upper and lower  $68 \%$ ($1 \sigma$) confidence limits for the medians. 
We thus find that $dEW$ is likely to be larger for systems with larger $EW$s, with the median of $dEW(1-2)$ being more than twice as large as that of $dEW(3-4)$. 
The same trend is also seen in the gradient of $EW$, $i.e.$, $dEW$ per unit separation, $dEW(i-j)/D_{\rm tra}(i-j)$, between the absorption lines in the same pairs of absorbers. 
The median  survival values of $dEW(1-2)/D_{\rm tra}(1-2)$, $dEW(2-3)/D_{\rm tra}(2-3)$, and $dEW(3-4)/D_{\rm tra}(3-4)$ with $68 \%$ ($1 \sigma$) confidence limits are estimated to be $0.024^{+0.002}_{-0.003}$ kpc$^{-1}$, $0.013^{+0.003}_{-0.002}$ kpc$^{-1}$, and $0.0096^{+0.0374}_{-0.0040}$ kpc$^{-1}$, respectively. 

These results suggest that, in a system composed of multiple absorbers, the absorbers giving rise to absorption lines with large $EW$s tend to have a relatively high degree of variation in absorption strength compared to those with small $EW$s. 
In the case where the absorbing systems are composed of numerous small clouds, the clouds exhibiting large $EW$s tend to have small covering factors and show a relatively high degree of incoherence producing the large $EW$ gradients. 
It should be noted, however,  that    
the magnitude correlations of the median survival values of both $EW$ and $dEW/D_{\rm tra}$ cannot be confirmed with the current statistics due to the small sample size of the multiple Mg\,{\sc ii} systems.
Further observations of the multiple Mg\,{\sc ii} systems are essential to achieve a conclusive statement about the spatial structure of gas and/or metals.

\subsection{Doublet Ratio}

The doublet ratio DR $(=W_{\rm MgII}^{0}(\lambda 2796)/W_{\rm MgII}^{0}(\lambda 2803))$ has been used as an alternative probe of the physical conditions of the absorbers giving rise to the Mg\,{\sc ii} absorption lines. 
Here we draw a detailed comparison of  DRs for each absorber in the separate sightlines.  
The doublet ratio reflects the degree of saturation in the Mg\,{\sc ii} doublet lines, which varies from 2 for unsaturated lines to $\sim 1$ for lines on the saturated part of the curve of growth \citep[e.g.][]{Stromgren48}. 

\begin{figure}[h]
\begin{center}
%\includegraphics[width=8cm]{/Users/okosh/Documents/data/DR_EW.pdf}
%[two column]\includegraphics[width=10cm]{fig13.pdf}
\includegraphics[width=8.5cm]{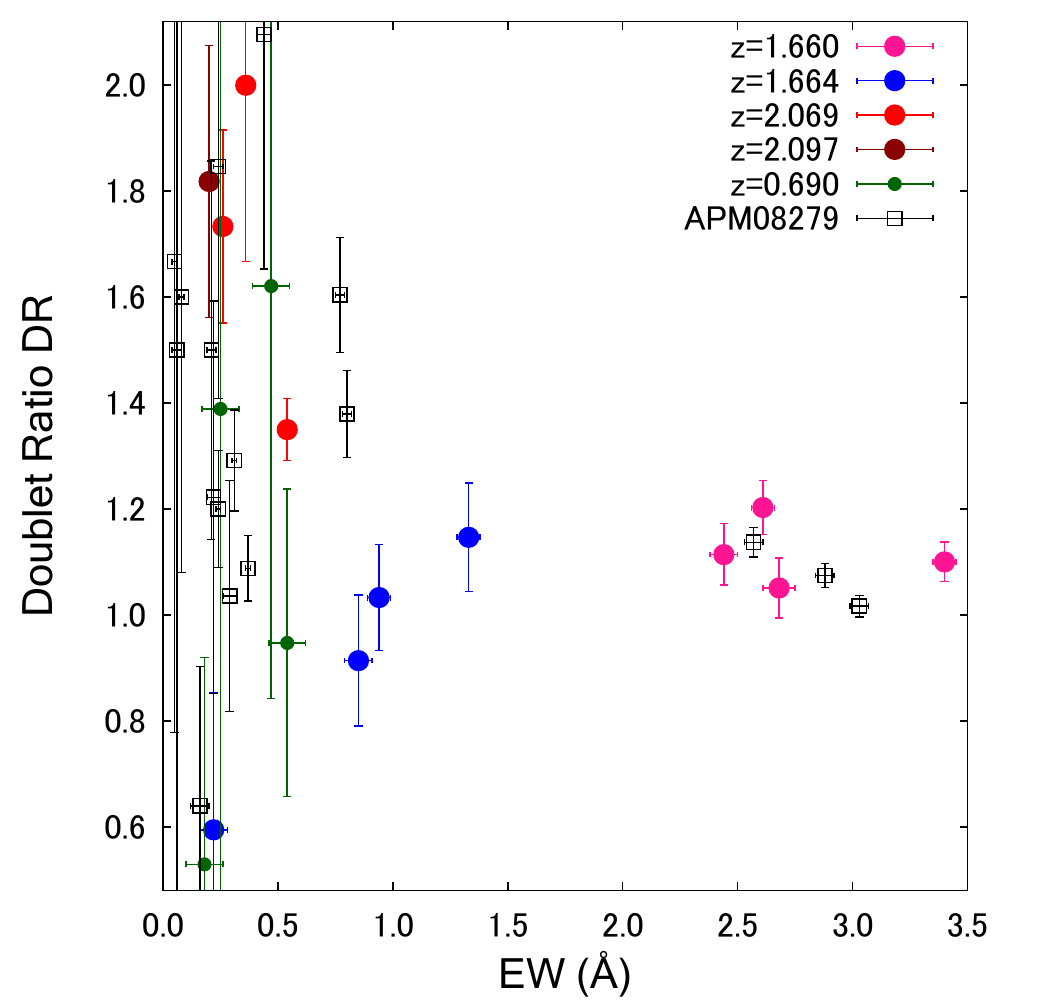}
\end{center}
\caption{The DRs as a function of rest equivalent width ($EW$) at $\lambda=2796$ ${\rm \AA}$ 
for the Mg\,{\sc ii} doublets in lines of sight toward quadruply/triply lensed quasars H1413+1143  (filled-circles) at 
$z_{\rm abs}=1.660$ (pink), $1.664$ (blue), $2.069$ (red), $2.097$ (dark red), $0.609$ (green), and APM08279+5255 (black open-squares) at $1 < z_{\rm abs} < 2 $ \citep{Ellison04}. 
\label{DR_EW}}
\end{figure}

In Table $\ref{table_DR}$, we show DRs of the Mg\,{\sc ii} doublets at $z_{\rm abs}=1.660$ (No.1a), $1.664$ (No.1b),  
$2.069$ (Nos.4-5), and $2.097$ (Nos.6-7) in the A/B/C/D spectra toward quasar H1413+1143. 
In Figure $\ref{DR_EW}$, we present the DRs as a function of rest EW at $\lambda=2796$ ${\rm \AA}$ for the Mg\,{\sc ii} doublets at $z_{\rm abs}=1.660$ (pink), $1.664$ (blue), $2.069$ (red), $2.097$ (dark red), and $0.609$ (green) \citep{Monier98}. 
To make a comparison to the DRs of Mg\,{\sc ii} doublets that have redshifts similar to the ones ($z_{\rm abs} > 1$) in our measurements,   
the $DRs$ of Mg\,{\sc ii} doublets at $z_{\rm abs} \sim 1-2$ in the spectra of the triply lensed quasar 
APM08279+5255 (black open-squares)\citep{Ellison04} are also plotted.
It is obvious that $DR$ values of the doublets at $z_{\rm abs}=1.660$ and $1.664$ (No.1a $\&$ No.1b) are relatively small ($\sim 1$)  
rather than the DR values of $\sim 2$ for the two doublets at $z_{\rm abs}=2.069$ and $2.097$ (Nos.4-5 $\&$ 6-7).  
The rest EWs of the former doublets at 
$z_{\rm abs}=1.660$ and $1.664$ (No.1a $\&$ 1b), $W_{\rm MgII}^{0}(\lambda 2796) >1$ ${\rm \AA}$, are relatively larger than those of the latter at $z_{\rm abs}=2.069$ and $2.097$, $W_{\rm MgII}^{0}(\lambda 2796) < 0.5$ ${\rm \AA}$. 
This suggests that the strong Mg\,{\sc ii} doublets likely exhibit the small DRs. 
The results are in good agreement with the values of DRs from a sample of Mg\,{\sc ii} doublets in the spectra of a triply lensed quasar 
APM08279+5255 (black open-squares)\citep{Ellison04}.

Our DR measurements for the Mg\,{\sc ii} doublets in the multiple lines of sight toward quadruply/triply lensed quasars are likely to exhibit an anticorrelation between DR and $W_{\rm MgII}^{0}(\lambda 2796)$,  which is also shown for the Mg\,{\sc ii} doublets in lines of sight toward single quasars \citep[e.g.][]{Lanzetta87, Steidel92, Zhu13}. 
We calculated the Spearman rank coefficient between the DR and $W_{\rm MgII}^{0}(\lambda 2796)$ for the sample presented in Figure $\ref{DR_EW}$. The sample gives the coefficient $\rho$ $=$ $-0.296$ and the $p$-value, $p$ $=0.094$, 
which does not rule out the null hypothesis that there is no correlation between the two quantities. 
However, for the samples with $1<$ DR $<2$, the coefficient $\rho$ $=$ $-0.650$ and $p$ $=2.4\times10^{-4}$, ruling out the null hypothesis. 
Since the sample size is still small,  more DR measurements using Mg\,{\sc ii} doublets in the spectra of quadruply/triply lensed quasars are required to reveal the correlation between the two quantities of the Mg\,{\sc ii} doublets at $z_{\rm abs} \sim 1-2$ in the spectra of the quadruply/triply lensed quasar. 

% Fig 13

Previously, detections of Mg\,{\sc ii} doublets indicate that strong Mg\,{\sc ii} systems likely give rise to Mg\,{\sc i} $\lambda 2853$ absorption lines \citep[e.g.][]{Steidel92}. 
In our sample, for the two strong Mg\,{\sc ii} systems at $z_{\rm abs}=1.660$ and $1.664$, Mg\,{\sc i} absorption lines (No.2 $\&$ 3) at $z_{\rm abs}=1.659$ and $1.667$ are detected in {\it all} the four separate spectra, which tends to occur in the Mg\,{\sc ii} systems giving rise to saturated doublets with small $DRs$ ($\sim 1$). 
By contrast, {\it no Mg\,{\sc i} absorption is detected} in the two weak Mg\,{\sc ii} systems at $z_{\rm abs}=2.069$ and $2.097$  
which exhibit the large $DRs$ ($\sim 2$).  
This suggests that the physical properties of the absorbers with saturated Mg\,{\sc ii} doublets are probably  similar to each other between the separation of sightlines on the scales of $6-12$ kpc.

\subsection{Gravitationally lensing galaxy}

Gravitational lensing provides valuable opportunities for investigating the mass distribution of the lens and the structure of the light source. The gravitationally lensed quasar H1413+1143 \citep[e.g.][]{Magain88} offers the four lensed images that display an almost symmetric configuration within $0.7$ arcsec of the image center. 
The clear images are expected to be a probe of the properties of the lensing body, but the nature, such as the lens redshift $z_{\rm lens}$, remains unclear. 
Detailed studies of the optical/infrared/radio images and the spectra have revealed the properties of the possible lensing galaxies, which are diffuse and faint and may belong to a galaxy cluster/group at $z > 1$ \citep[e.g.][]{Turnshek97, Kneib98a, Kneib98b, MacLeod09}. 

The time delay between the lensed images, which depends on the gravitational potential of the lensing bodies and the geometrical properties of the universe at large distances, is well known to be an important feature of the gravitational lensing as a diagnostic for the properties of the lensing body \citep[e.g.][]{Refsdal64a, Refsdal64b}. 
For the lensed quasar H1413+1143, long-term photometric observations have been carried out to measure the time delays between the four lensed images \citep[e.g.][]{Kayser90, Goicoechea10, Akhunov17}. 
The measurements of the time delays with the Sloan Digital-Sky Survey improved the lens model for quasar H1413+1143 and provided estimations of possible redshifts of the lensing galaxy: $z_{\rm lens}=1.88_{-0.11}^{+0.09} $\citep{Goicoechea10} and $1.95_{-0.10}^{+0.06} $\citep{Akhunov17}. 
The time-delay measurements imply a possible redshift of the lensing galaxy:  $z_{\rm lens}$ $\sim1.9$. 
In an assumption that the lensing galaxies have galactic halos and/or the CGM where the gas clouds or galactic flows give rise to  absorption signatures in the spectra of the lensed quasar \citep[e.g.][]{Zahedy16}, Mg\,{\sc ii} absorption lines are expected to be detected  at the possible redshift of the lensing galaxy, $z_{\rm lens} \sim 1.9$, where the wavelength coverage of the Kyoto 3D\,{\sc ii} allows the detection of the Mg\,{\sc ii} absorption doublets. 
We find no signature of the Mg\,{\sc ii} absorption lines at $z \sim 1.9$ in the spatially resolved four spectra with the Kyoto 3D\,{\sc ii} (Figure $\ref{spec_figure_new}$). 
It should be noted that, for the lensing galaxy at $z_{\rm abs}$ $\sim$ $1.9$, we place a $3\sigma$ upper limit on the rest EW as $\sim$  $0.03$ ${\rm \AA}$, assuming its line with of $300$ km s$^{-1}$, as we did for the other undetected Mg\,{\sc ii} absorption lines in Section 3.
Under our detection limit, this leads to an implication that the multiple lines of sight do not intersect with Mg\,{\sc ii} absorbers embedded in the halo and/or the CGM of the possible lensing galaxy at $z \sim 1.9$ which redshifts are estimated 
on the basis of the measurements of the time delays.

\section{Conclusions}

We have investigated the properties of multiple and coincident Mg\,{\sc ii} absorption systems 
in the {\it spatially resolved four} spectra toward the four images A/B/C/D of the quadruply gravitationally lensed quasar H1413+1143 on the basis of the high spatial resolution and high-S/N spectroscopy with an optical multi-mode spectrograph, the Kyoto 3D\,{\sc ii}, on board the Subaru telescope.
For the quadruply gravitationally lensed quasar H1413+1143, there have been the previous measurements of H\,{\sc i} and metal absorption strengths, except for Mg\,{\sc ii}. 
We find multiple Mg\,{\sc ii} absorption systems at redshifts $z=1.66$, $2.069$, and $2.097$ in the spatially resolved Kyoto 3D\,{\sc ii} spectra. 
Here we present the first measurement of differences in the Mg\,{\sc ii} absorption strength of the multiple intervening absorbers. 
The main conclusions are summarized as follows.\\

1. The Mg\,{\sc ii} absorption systems at $z_{\rm abs} = 1.66$ give rise to a strong Mg\,{\sc ii} absorption line, which consists of two components at $z_{\rm abs} = 1.660$ and $1.664$, in {\it all} four spatially resolved spectra toward images A/B/C/D (the A/B/C/D spectra) of quasar H1413+1143. 
In the four spectra, for the Mg\,{\sc ii} absorption lines at $z_{\rm abs}=1.66$, the rest equivalent width of the strong (blended) Mg\,{\sc ii} absorption line in the A spectrum is the largest, whereas that in the D spectrum is the smallest. 
The rest equivalent widths change by factors of up to $\sim 1.4$. 
For the Mg\,{\sc ii} absorption component at $z_{\rm abs} = 1.660$, the rest equivalent widths in the four spectra change by factors of up to $ \sim 1.5$: A $>$ D $>$ B $>$ C. 
By contrast, for the Mg\,{\sc ii} absorption component at $z_{\rm abs} = 1.664$, the rest equivalent widths change by factors of up to $ \sim 6$, B $>$ A $>$ C $>$ D which is consistent with the variation of H\,{\sc i} column density, B $>$ A $>$ C $>$ D, in a DLA system identified 
at $z_{\rm abs}=1.662$ in the HST observation.  
The variations in the equivalent width of the Mg\,{\sc ii} lines at $z_{\rm abs}=1.66$ in the spectra are also in agreement with those of the other metal absorption lines (e.g., C\,{\sc ii}, Si\,{\sc ii}, Al\,{\sc ii}, Fe\,{\sc ii}, C\,{\sc iv}, Si\,{\sc iv}) in the previous observations \citep{Monier98}, except for the small equivalent widths of the Mg\,{\sc ii} lines at $z_{\rm abs}=1.664$  in the D spectrum. \\

2. The absorption systems at $z_{\rm abs} = 2.069$ give rise to strong Mg\,{\sc ii} absorption lines in the three A/B/C spectra,   
whereas no Mg\,{\sc ii} absorption is detected in the D spectrum. 
The rest equivalent widths in the three spectra change by factors of up to $\sim 2$: A $>$ C $>$ B. 
The ratios between the equivalent widths of the Mg\,{\sc ii} lines in the three spectra are in agreement with those between the equivalent widths of the other metal lines (Si\,{\sc iv} and C\,{\sc iv}). \\

3. The absorption systems at $z_{\rm abs} = 2.097$ give rise to a Mg\,{\sc ii} absorption line in the B spectrum,  
whereas no Mg\,{\sc ii} absorption is detected in the other three A/C/D spectra.  
This system is also identified as a sub-DLA system \citep{Monier09}. 
In the four spectra, the sub-DLA system gives rise to the strongest H\,{\sc i} absorption lines with H\,{\sc i} column density  $N_{\rm HI} \sim 5 \times 10^{19}$ cm$^{-2}$ in the B spectrum. 
This suggests that the sub-DLA system is associated with the Mg\,{\sc ii} absorption system.  \\ 

4. The multiple Mg\,{\sc ii} systems in the separate spectra toward the lensed or pair quasars, including our measurements for those in the spectra of the lensed quasar H1413+1143 tend to exhibit large fractional difference $dEWs$ for the systems with large separations between lines of sight.
There is no evidence for a relationship between the fractional difference in equivalent width and the equivalent widths of the absorption lines. 
Focusing on the multiple Mg\,{\sc ii} systems in the spectra of {\it quadruply gravitationally lensed quasars}, 
for the individual system that is composed of the multiple absorbers, the absorbers giving rise to absorption lines with large equivalent widths tend to have relatively large fractional differences in equivalent width, which is suggestive of a high degree of variation in absorption strength, rather than the absorbers with small equivalent widths. \\

5. The doublet ratios (DRs) of the equivalent width of the Mg\,{\sc ii} absorption line at $\lambda = 2796$ ${\rm \AA}$ to that at $\lambda = 2803$ ${\rm \AA}$ (= $W_{\rm MgII}^{0}(\lambda 2796)/W_{\rm MgII}^{0}(\lambda 2803)$) are close to $\sim 1$ 
for the strong doublets, with $W_{\rm MgII}^{0}(\lambda 2796) > 1$ ${\rm \AA}$ at $z_{\rm abs}=1.660$ and $1.664$ in the spectra of quasar H1413+1143. 
The weak doublets with $W_{\rm MgII}^{0}(\lambda 2796) < 0.5$ ${\rm \AA}$ at $z_{\rm abs}=2.069$ and $2.097$ have the ratios DRs of $\sim 2$. 
The DRs of the multiple Mg\,{\sc ii} systems in the spectra of the quasar H1413+1143 likely exhibit an anticorrelation 
between the DR and $W_{\rm MgII}^{0}(\lambda 2796)$. 
The DR values of the Mg\,{\sc ii} doublets in our measurements are in good agreement with those of the Mg\,{\sc ii} doublets at $z_{\rm abs} > 1$ in the previous measurements. 
Mg\,{\sc i} $\lambda 2853$ absorption lines at $z_{\rm abs}=1.659$ and $1.667$ are also identified in {\it all} four separate spectra, which tends to occur in the strong Mg\,{\sc ii} systems with the {\it small} DRs ($\sim 1$) at $z_{\rm abs}=1.66$.  
By contrast, no Mg\,{\sc i} absorption is detected in the two weak Mg\,{\sc ii} systems with the {\it large} DRs ($\sim 2$) at $z_{\rm abs}=2.069$ and $2.097$. \\

6. We find no signature of the Mg\,{\sc ii} absorption lines at $z \sim 1.9$ where the possible lensing galaxy is expected to be detected on the basis of the measurements of the time delays between the lensed images. \\

\acknowledgments

We thank S. Koyamada for valuable discussions of this study and the anonymous referee for a careful reading of this manuscript and suggestions that improved the clarity of this presentation. 
This work has been supported by a grant from JSPS Grant Numbers JP16K05299(KO) and JP21K13956(DK). 
We extend thanks to the indigenous Hawaiian community and are grateful to have the opportunity to carry out the observations from Maunakea.


%%%%%%%%%%%%%%%%%%%%%%%%%%%%%%%%%%%%%%%%%%%%%%%%%%%%%%%%%\bibliography{sample63}{}
%%%%%%%%%%%%%%%%%%%%%%%%%%%%%%%%%%%%%%%%%%%%%%%%%%%%%%%%%\bibliographystyle{aasjournal}



%%%%%%%%  Figure Captions  %%%%%%%%%%%%%%%%%

%%%%%%%%%%%%%
\newpage
%%%%%%%%%%%%%

\begin{table}
\begin{center}
\caption{Rest Equivalent Widths of Multiple Absorption Systems}
\label{table_EW}
\begin{tabular}{cc|cc|cc|cc|cc} 
\hline
&  & A & & B & & C & & D & \\ 
\hline
No & ion & $z_{\rm abs}$ & $W_{\rm rest}$[{\rm \AA}] & $z_{\rm abs}$ & $W_{\rm rest}$[{\rm \AA}] & $z_{\rm abs}$ & 
$W_{\rm rest}$[{\rm \AA}] & $z_{\rm abs}$ & $W_{\rm rest}$[{\rm \AA}] \\
\hline
1 & Mg\,{\sc ii}$\lambda \lambda$2796,2803 & 1.66 & 8.21$\pm$0.02 & 1.66 & 7.09$\pm$0.02 & 1.66 & 6.23$\pm$0.02 & 1.66 & 5.73$\pm$0.03 \\

1a & Mg\,{\sc ii}$\lambda$2796 & 1.660 & 3.40$\pm$0.05 & 1.660 & 2.61$\pm$0.05 & 1.660 & 2.44$\pm$0.06 & 1.660 & 2.68$\pm$0.07 \\

 & Mg\,{\sc ii}$\lambda$2803 & & 3.09$\pm$0.06 &  & 2.17$\pm$0.05 & & 2.19$\pm$0.06 & & 2.55$\pm$0.07 \\

1b & Mg\,{\sc ii}$\lambda$2796 & 1.664 & 0.94$\pm$0.05 & 1.664 & 1.33$\pm$0.05 & 1.665 & 0.85$\pm$0.06 & 1.665 & 0.22$\pm$0.06 \\

 & Mg\,{\sc ii}$\lambda$2803 & & 0.91$\pm$0.04 & & 1.16$\pm$0.06 & & 0.93$\pm$0.06 & & 0.37$\pm$0.06 \\

2 & Mg\,{\sc i}$\lambda$2853 & 1.659 & 0.44$\pm$0.05 & 1.659 & 0.56$\pm$0.06 & 1.659 & 0.27$\pm$0.05 & 1.659 & 0.43$\pm$0.06  \\

3 & Mg\,{\sc i}$\lambda$2853 & 1.667 & 0.25$\pm$0.02 & 1.668 & 0.19$\pm$0.02 & 1.667 & 0.32$\pm$0.03 & 1.668 & 0.26$\pm$0.02  \\ 
\hline 

4 & Mg\,{\sc ii}$\lambda$2796 & 2.069 & 0.54$\pm$0.01 & 2.069 & 0.26$\pm$0.01 &  2.069 &  0.36$\pm$0.02 & --- & $<$0.05$^{b}$ \\
5 & Mg\,{\sc ii}$\lambda$2803 &  & 0.40$\pm$0.01 & & 0.15$\pm$0.01 & & 
0.18$\pm$0.02 & & $<$0.05$^{b}$ \\
\hline 

6 & Mg\,{\sc ii}$\lambda$2796 & --- & $<$0.03$^{b}$ & 2.097  & 0.20$\pm$0.01 & --- &  $<$0.03$^{b}$ & --- & $<$0.05$^{b}$ \\
7 & Mg\,{\sc ii}$\lambda$2803 & & $<$0.03$^{b}$ & & 0.11$\pm$0.01 & & $<$0.03$^{b}$ & & $<$0.05$^{b}$ \\
\hline
\hline
& & $z_{\rm abs}$ & $N_{\rm HI}$[$10^{20}$cm$^{-2}$] & $z_{\rm abs}$ & $N_{\rm HI}$[$10^{20}$cm$^{-2}$] & $z_{\rm abs}$ & $N_{\rm HI}$[$10^{20}$cm$^{-2}$] & $z_{\rm abs}$ & $N_{\rm HI}$[$10^{20}$cm$^{-2}$] \\
\hline
& H\,{\sc i}$\lambda$1216$^{a}$ & 1.662 & 1.5$\pm$0.45 & 1.662 & 6.0$\pm$1.8 & 1.662 & 0.60$\pm$0.18 & 1.662 & 0.30$\pm$0.09 \\ 
\hline
\end{tabular}
\end{center}
\tablecomments{(a) \citet{Monier09}; HST-FOS. 
(b) 3$\sigma$ detection limit assuming a line width of $300$ km s$^{-1}$, a typical value of the other detected lines with a single component.}
\end{table}

\newpage

\begin{table}
\begin{center}
\caption{Multiple Mg\,{\sc ii} absorptions at $\lambda2796$ ${\rm \AA}$ in lines of sight toward Quadruply or Triply Lensed Quasars}
\label{table_QSO}
\begin{tabular}{cccccccc} \hline
QSO & $z_{\rm abs}$ & A [$\rm \AA$] & B [$\rm \AA$] &  C [$\rm \AA$] & D [$\rm \AA$] & 
$d_{\rm max}$/kpc $^{a}$& $d_{\rm min}$/kpc  $^{a}$\\
\hline
H1413+1143 $^{b}$ & 0.6089  &
0.25$\pm$0.09 & 0.19$\pm$0.09 & 0.55$\pm$0.09 & 0.48$\pm$0.08 
& 9.1 & 5.1 \\
& 1.660 &
3.40$\pm$0.05 & 2.61$\pm$0.05 & 2.44$\pm$0.06 & 2.68$\pm$0.07 
&  11.5 &  6.4 \\
& 1.664 &
0.94$\pm$0.05 & 1.33$\pm$0.05 & 0.85$\pm$0.06 & 0.22$\pm$0.06 
&  11.5 & 6.4  \\
& 2.069 &
0.54$\pm$0.01 & 0.26$\pm$0.01 & 0.36$\pm$0.03 & $<$0.05
&  7.6 & 4.2 \\
& 2.097 &
$<$0.03 & 0.20$\pm$0.01 & $<$0.03 & $<$0.05
&  7.1 & 3.9 \\
& & & & & & & \\
HE0435-1223 $^{c}$ & 0.4188 & 
1.1$\pm$0.1 & 1.9$\pm$0.1 & 1.5$\pm$0.1 & 1.0$\pm$0.1 
& 13.8 & 8.2 \\
& 0.7818 & 
0.51$\pm$0.03 & --- & 0.72$\pm$0.03 & 0.56$\pm$0.04 & 7.8 & 5.7 \\
& & & & & & & \\
J014710+463040 $^{d}$ & 0.577 & 
0.1111$\pm$0.0052 & 0.1138$\pm$0.0053 & 0.1502$\pm$0.0078 & 0.1163$\pm$0.0171 
& 22.1 & 8.3 \\ 
& 0.576 & --- & --- & --- & 0.1252$\pm$0.0170 & & \\
& 0.607 & 
0.4621$\pm$0.0050 & 0.2814$\pm$0.0063 & 0.3736$\pm$0.0072 & 0.2907$\pm$0.0208 
& 21.1 & 8.0 \\ 
& 0.758 & 
$<$0.0849 & $<$0.0792 & $<$0.1128 & 0.9831$\pm$0.0246 
& 16.8 & 6.4 \\ 
& & & & & & & \\
Q2237+0305 $^{e}$ & 0.566 & 
0.65$\pm$0.07 & 0.65$\pm$0.06 & 0.78$\pm$0.07 & --- 
& 0.5 & 0.4 \\
 & 0.827 & 0.05$\pm$0.06 & 0.07$\pm$0.01 & --- & --- & 0.3 & 0.3 \\
& & & & & & & \\
J1004+4112 $^{f}$ & 0.676 & 0.91$\pm$0.03 & $<$0.2 & $<$0.2 & 1.0$\pm$0.1 
& 103.0 & 26.3 \\
& 0.726 &
$<$0.1 & $<$0.2 & $<$0.3 & 2.6$\pm$0.1 
& 94.7 & 24.2 \\
& 0.749 &
$<$0.1 & $<$0.1 & $<$0.2 & 1.4$\pm$0.1 
& 90.7 & 23.2 \\
& 0.833 &
$<$0.1 & 0.8$\pm$0.1 & $<$0.2 & $<$0.4 
& 77.1 & 19.7 \\
& 1.022 &
$<$0.1 & 0.23$\pm$0.03 & $<$0.1 & $<$0.2 
& 52.2 & 13.3 \\
& 1.083 &
$<$0.1 & $<$0.2 & $<$0.2 & 1.5$\pm$0.1 
& 45.5 & 11.6 \\
& 1.226 &
$<$0.1 & $<$0.1 & $<$0.2 & 0.5$\pm$0.1 
& 31.9 & 8.1 \\
& 1.258 &
$<$0.1 & $<$0.1 & $<$0.2 & 0.9$\pm$0.1 
& 29.2 & 7.5 \\
& & & & & & & \\
APM08279+5255 $^{g}$ & 1.181 & 
2.57$\pm$0.04 & 3.03$\pm$0.04 & 2.28$\pm$0.04 & ---
& 2.7 & 1.0 \\
& 1.209 & 
0.05$\pm$0.01 & 0.06$\pm$0.02 & $<$0.03 & ---
& 2.6 & 1.0 \\
& 1.211 & 
0.37$\pm$0.01 & $<$0.03 & 0.16$\pm$0.04 & ---
& 2.6 & 1.0 \\
& 1.291 & 
0.08$\pm$0.01 & 0.03$\pm$0.01 & $<$0.03 & ---
& 2.4 & 0.9 \\
& 1.444 & 
0.04$\pm$0.01 & 0.04$\pm$0.01 & $<$0.03 & ---
& 2.0 & 0.8 \\
& 1.550 & 
0.31$\pm$0.01 & $<$0.02 & 0.29$\pm$0.03 & ---
& 1.8 & 0.7 \\
& 1.552 & 
0.24$\pm$0.01 & $<$0.02 & $<$0.03 & ---
& 1.8 & 0.7 \\
& 1.813 & 
0.80$\pm$0.02 & 0.77$\pm$0.02 & 0.44$\pm$0.03 & ---
& 1.4 & 0.5 \\
& 2.041 & 
0.21$\pm$0.02 & 0.24$\pm$0.02 & 0.22$\pm$0.03 & ---
& 1.1 & 0.4 \\
& 2.066 & 
0.31$\pm$0.04 & 0.45$\pm$0.04 & 0.38$\pm$0.04 & ---
& 1.0 & 0.4 \\
\hline
\end{tabular}
\end{center}
\tablecomments{(a) $d_{\rm max}$ and $d_{\rm min}$ present the largest and smallest proper separations between the images in the absorbers at the absorption redshift, respectively.
(b)This study ($z_{\rm abs}=1.660, 1.664, 2.069$, and $2.097$) $\&$ \citet{Monier98} ($z_{\rm abs}=0.6089$); $z_{\rm source}=2.54$, and $z_{\rm lens}=1.90$ are adopted. 
(c)\citet{Chen14}; $z_{\rm source}=1.689$, $z_{\rm lens}=0.4546$. 
(d)\citet{Rubin18b}; $z_{\rm source}=2.377$, $z_{\rm lens}=0.5768$. 
(e)\citet{Rauch02}; $z_{\rm source}=1.69$, $z_{\rm lens}=0.039$. The Mg\,{\sc ii} line at $z=0.827$ is negligible in the C spectrum.  
(f)\citet{Oguri04}; $z_{\rm source}=1.734$, $z_{\rm lens}=0.68$. 
(g)\citet{Ellison04}; $z_{\rm source}=3.911$, $z_{\rm lens}=1.062$. }
\end{table}

\newpage

\begin{table}
\begin{center}
\caption{Doublet Ratios of Rest Equivalent Widths}
\label{table_DR}
\begin{tabular}{c|cc|cc|cc|cc} 
\hline
& A & & B & & C & & D &  \\ 
\hline
No  & $W(\lambda2796)$[{\rm \AA}] & $\frac{\displaystyle W(\lambda2796)}{\displaystyle W(\lambda2803)}$ & 
$W(\lambda2796)$[{\rm \AA}] & $\frac{\displaystyle W(\lambda2796)}{\displaystyle W(\lambda2803)}$ & 
$W(\lambda2796)$[{\rm \AA}] & $\frac{\displaystyle W(\lambda2796)}{\displaystyle W(\lambda2803)}$ & 
$W(\lambda2796)$[{\rm \AA}] & $\frac{\displaystyle W(\lambda2796)}{\displaystyle W(\lambda2803)}$ \\
\hline
1a & 3.40$\pm$0.05 & 1.10$\pm$0.04 & 2.61$\pm$0.05 & 1.20$\pm$0.05 & 2.44$\pm$0.06 & 1.11$\pm$0.06 & 2.68$\pm$0.07 & 1.05$\pm$0.06 \\

1b & 0.94$\pm$0.05 & 1.03$\pm$0.10 & 1.33$\pm$0.05 & 1.15$\pm$0.10 & 0.85$\pm$0.06 & 0.91$\pm$0.12 & 0.22$\pm$0.06 & 0.59$\pm$0.26 \\

4-5 & 0.54$\pm$0.01 & 1.35$\pm$0.06 & 0.26$\pm$0.01 &  1.72$\pm$0.18 &  0.36$\pm$0.02 & 2.00$\pm$0.33 & $<$0.05 & --- \\
6-7 & $<$0.03 & --- & 0.20$\pm$0.01 & 1.82$\pm$0.26 &  $<$0.03 & --- & $<$0.05 & --- \\
\hline
\end{tabular}
\end{center}
\end{table}

%%%%%%%%%%%
%\newpage
%%%%%%%%%%%

\end{document}